\documentclass[a4paper,11pt]{article}
\usepackage{jheppub} 
\usepackage{lineno}
\usepackage[normalem]{ulem}
\usepackage{bm}
\usepackage{soul}
\usepackage{amsmath}

\title{\boldmath A Predictive Non-Holomorphic Modular $A_4$ Linear Seesaw Framework Testable at DUNE}

\author[a]{Rudra Majhi,}
\author[b]{Mitesh Kumar Behera,}
\author[c]{Rukmani Mohanta}

\affiliation[a]{Department of Physics, Nabarangpur College, Nabarangpur, Odisha, 764059, India}
\affiliation[b]{Department of Physics, School of Advanced Sciences, Vellore Institute of Technology, Tiruvalam Rd, Katpadi, Vellore, Tamil Nadu 632014, India}
\affiliation[c]{School of Physics, University of Hyderabad, Hyderabad-500046, India}

\emailAdd{rudra.majhi95@gmail.com}
\emailAdd{miteshbehera1304@gmail.com}
\emailAdd{rmsp@uohyd.ac.in}

\abstract{
We study a realization of neutrino masses and mixing phenomena within a linear seesaw mechanism based on non-holomorphic modular $A_4$ symmetry, which extends modular-invariant flavor models beyond the conventional holomorphic framework. The model is constructed in a non-supersymmetric setting and involves six heavy $SU(2)_L$ singlet fermions, $N_{Ri}$ and $S_{Li}$, together with a single flavon field, thereby significantly reducing the field content  compared to conventional $A_4$ flavor models that typically require multiple flavon fields as well as supersymmetric (holomorphic) modular frameworks involving additional superfields. The modular transformation properties of the Yukawa couplings under $A_4$ symmetry lead to a highly constrained neutrino mass matrix with a distinctive flavor structure. After presenting the general theoretical framework, we perform a systematic numerical analysis of neutrino phenomenology by restricting the modulus parameter $\tau$ to the fundamental domain and scanning the allowed parameter space. We identify regions consistent with current neutrino oscillation data at the $3\sigma$ level and obtain predictions for currently unknown observables, including the absolute neutrino mass scale and leptonic CP-violating phases. We further examine the implications for neutrinoless double beta decay, highlighting testable signatures in the upcoming precision oscillation as well as rare-process experiments. These results demonstrate the phenomenological viability and predictive power of non-holomorphic modular symmetry in linear seesaw neutrino mass models.
}

\begin{document}
\maketitle
\flushbottom

\section{Introduction}
\label{sec:intro}

The Standard Model (SM) falls short in explaining the observed properties of neutrinos. While it predicts neutrinos to be massless, experimental results show that they have tiny but non-zero masses \cite{Faessler:2020sgs,Aker:2019uuj,Pontecorvo:1967fh,Feruglio:2017ieh}, as confirmed by neutrino oscillation data. Neutrino oscillations are now firmly established and provide strong evidence that neutrinos mix with each other, with at least two of them having non-zero masses \cite{Tanabashi:2018oca}. In the SM, neutrinos lack right-handed (RH) counterparts, which prevents them from acquiring Dirac masses like other charged fermions. However, the dimension-five Weinberg operator \cite{Weinberg:1980bf,Weinberg:1979sa,Wilczek:1979hc} offers a way to generate neutrino masses. Still, the fundamental origin and flavor structure of this operator remain open questions.

Therefore, it becomes essential to explore scenarios beyond the Standard Model (BSM) to account for the non-zero masses of neutrinos. Several models have been proposed in the literature to explain the results from neutrino oscillation experiments. These include the well-known seesaw mechanisms \cite{Minkowski:1977sc,Mohapatra:1979ia,GellMann:1980vs}, radiative mass generation models \cite{Zee:1980ai,Babu:1988ki}, and frameworks involving extra dimensions \cite{ArkaniHamed:1998vp}, among others. A common feature of many such BSM approaches is the introduction of SM gauge singlet fermions, often identified as right-handed neutrinos, that couple to active neutrinos via Yukawa interactions. Their masses and couplings can vary across many orders of magnitude, leading to a wide range of possible phenomenological implications. For instance, in the canonical seesaw mechanism, explaining eV-scale light neutrino masses typically requires right-handed neutrino masses around $10^{15}$ GeV, which is far beyond the reach of present or future experiments. In contrast, low-scale variants such as the inverse seesaw \cite{Mohapatra:1986bd,GonzalezGarcia:1988rw,CarcamoHernandez:2019pmy}, linear seesaw \cite{Malinsky:2005bi}, and extended seesaw \cite{Mohapatra:2005wg} allow the heavy neutrinos to lie at the TeV scale, making them accessible for experimental verification. \vspace{1mm}

The non-Abelian discrete flavor symmetry group $A_4$ has long been considered a compelling framework for explaining the structure of the neutrino mass matrix~\cite{Ma:2001dn, Ishimori:2012zz}. In its simplest realization, however, it predicts a vanishing reactor mixing angle $\theta_{13}$, which contradicts the experimental observation. To reconcile this, additional flavon fields,  Standard Model singlet scalars, transforming non-trivially  under $A_4$ are typically introduced. Their vacuum alignments break the flavor symmetry spontaneously, generating a non-zero $\theta_{13}$ and reproducing realistic mixing patterns~\cite{Pakvasa:1977in}. Despite these successes, such constructions often rely on multiple flavons and higher-dimensional operators, which can obscure predictability and restrict the flavor symmetry’s role to constrain the mixing angles rather than determining the full neutrino mass structure.

A more predictive and elegant alternative arises in the \textit{modular invariance approach} originally proposed by Feruglio~\cite{Feruglio:2017spp} and thereafter followed by myriad literature \cite{Abbas:2020qzc,King:2020qaj,Wang:2019xbo, Lu:2019vgm,Nomura:2019xsb,Asaka:2019vev,Penedo:2018nmg,Liu:2020akv,Kobayashi:2019xvz,Novichkov:2018ovf,Novichkov:2020eep,Gui-JunDing:2019wap,Ding:2019zxk,Novichkov:2018nkm,Nomura:2019lnr,Ma:2015fpa,Mishra:2019oqq, Behera:2020sfe, Behera:2020lpd, Behera:2021eut, Mishra:2022egy, Mishra:2023ekx, Kumar:2023moh, Behera:2024ark, Behera:2025tpj}. Here, Yukawa couplings are promoted to \textit{modular forms}, transforming under the finite modular groups $\Gamma(N)$ or $\Gamma'(N)$. This framework naturally eliminates the need for flavon fields while maintaining a high degree of theoretical control. Traditionally studied in supersymmetric (SUSY) settings, modular symmetry ensures holomorphicity of the superpotential. However, with the absence of low-energy SUSY signals and growing experimental constraints, recent developments have extended this idea to \textit{non-holomorphic modular symmetries} originally given in Ref. \cite{Qu:2024rns}, vastly enriching the theoretical landscape.

In particular, 
\textit{polyharmonic Maa\ss{} modular forms}, which contain both holomorphic and non-holomorphic components, provide a broader and more flexible structure for model building. These forms satisfy Laplacian conditions and decompose into irreducible representations of $\Gamma(N)$ or $\Gamma'(N)$, allowing for richer flavor dynamics without invoking additional scalar fields. This expanded framework offers distinct advantages: it maintains modular predictability, accommodates radiative and non-supersymmetric effects, and provides new avenues for generating viable neutrino masses and mixings as shown in Refs.\cite{Tapender:2026ets, Nasri:2026nbf, Priya:2025wdm, Nomura:2024atp, Nomura:2024vzw, Nomura:2024nwh, 
Nomura:2025ovm, Nomura:2025raf, Nomura:2025bph, 
Kang:2024jnp, Ding:2024inn, Li:2024svh, Okada:2025jjo, Kobayashi:2025hnc, Loualidi:2025tgw, Zhang:2025dsa, Nomura:2024ctl, Abbas:2025nlv, Li:2025kcr, Dey:2025zld,Nanda:2025lem,Kumar:2025bfe}. Thus, non-holomorphic modular invariance not only overcomes the limitations of discrete flavor models but also establishes a versatile and predictive framework for neutrino physics that does not involve supersymmetry. 

The structure of this paper is as follows. In Sec. \ref{sec:non-holomorphic}, we outline the brief insights into non-holomorphic modular symmetry. We then provide a discussion of the light neutrino masses and mixing in this model framework in Sec. \ref{sec:linear}, and simulation details are discussed in Sec. \ref{simulation}. Further in Sec. \ref{sec:results}, a numerical correlational study is established between observables of the neutrino sector and model input parameters. In Section \ref{sec: summary}, we conclude our results.

\section{Non-Holomorphic Modular Symmetry}
\label{sec:non-holomorphic}

\noindent
Modular symmetry arises geometrically as the symmetry of a two-dimensional
torus. A torus $T^2$ is constructed by identifying points in the Euclidean
plane $\mathbb{R}^2$ under translations of a lattice $\Lambda$, such that
$T^2 = \mathbb{R}^2/\Lambda$. In the complex representation, the lattice is
spanned by two basis vectors, and the ratio of these vectors determines the
complex structure of the torus~\cite{Ishimori:2010au,Chauhan:2023faf}. This
ratio is encoded in the complex modulus $\tau = x + i y$, which transforms
under modular symmetry as
\begin{equation}
\tau \;\longrightarrow\; \gamma \tau
= \frac{a\tau + b}{c\tau + d}\;,
\qquad
\gamma =
\begin{pmatrix}
a & b \\ c & d
\end{pmatrix},
\end{equation}
where $a,b,c,d \in \mathbb{Z}$ satisfying $ad - bc = 1$. These transformations
generate the modular group $SL(2,\mathbb{Z})$, representing the symmetry of
the torus. Its finite quotients $\Gamma_N$ for $N=2,3,4,5$ are isomorphic to
the discrete groups  $S_3$ \cite{Kobayashi:2018wkl, Okada:2019xqk, Mishra:2020gxg,Kang:2026osw, Meloni:2023aru, Marciano:2024nwm, Belfkir:2024wdn}, $A_4$ \cite{Nomura:2023usj, RickyDevi:2024ijc,Gogoi:2023jzl,Pathak:2024sei,Nomura:2024vus,Kashav:2021zir, Kashav:2022kpk, Kobayashi:2019gtp, Nomura:2023kwz, Kim:2023jto, Devi:2023vpe, Dasgupta:2021ggp, CentellesChulia:2023osj}, $S_4$~\cite{deMedeirosVarzielas:2023crv, Ding:2021zbg, King:2019vhv}, and $A_5$~\cite{deMedeirosVarzielas:2022ihu, Yao:2020zml}, respectively.

\medskip
\noindent
Modular invariance offers a predictive approach to flavor model building,
providing an elegant explanation of fermion mass hierarchies and mixing
structures. In $\mathcal{N}=1$ supersymmetric (SUSY) constructions, modular
symmetry has been successfully implemented to reproduce realistic lepton
masses and mixing patterns with a minimal set of free parameters~\cite{Feruglio:2017spp}.
In such setups, matter superfields transform under representations of
$SL(2,\mathbb{Z})$ (or its projective counterpart $PSL(2,\mathbb{Z})$), and
the Yukawa couplings emerge as holomorphic modular forms of the modulus
$\tau$, defined over the upper half-plane.

\medskip
\noindent
While SUSY frameworks naturally ensure holomorphic Yukawa couplings, strong
constraints from flavor observables, electroweak precision measurements, and
astrophysical data severely limit the viable SUSY parameter space. Moreover,
the absence of experimental evidence for low-energy supersymmetry motivates
the exploration of non-SUSY realizations of modular flavor symmetry. It has been emphasized that the modular-form structure of Yukawa couplings is
not intrinsically tied to supersymmetry; rather, it depends on the details of
the compactification geometry. Recently, non-SUSY implementations of modular
flavor symmetry have been developed using automorphic forms, where the
holomorphicity condition is replaced by a Laplacian constraint. For a single
complex modulus, these automorphic functions correspond to \textit{harmonic
Maa\ss{} forms}, which are non-holomorphic modular functions satisfying the
Laplace equation. Consequently, modular flavor symmetry can be consistently
realized even in the absence of SUSY~\cite{Qu:2024rns}.

\medskip
\noindent
Within this approach, the relevant mathematical objects are
\textit{polyharmonic Maa\ss{} forms} associated with the modulus $\tau$. For a
fixed level $N$, the finite modular group $\Gamma'_N$ (or equivalently
$\Gamma_N$) coincides with that of the holomorphic case. Generic matter
fields $\psi_i$ and $\psi_i^c$ transform under the modular symmetry with
modular weights $-k_\psi$ and $-k_{\psi^c}$, and according to irreducible
representations $\rho_\psi$ and $\rho_{\psi^c}$ of $\Gamma'_N$ (or
$\Gamma_N$), respectively:

\begin{equation}
   \tau \to \gamma \tau = \frac{a\tau + b}{c\tau + d}, \quad 
   \gamma = 
   \begin{pmatrix} a & b \\ c & d \end{pmatrix} \in SL(2, \mathbb{Z}),
\end{equation}
\begin{equation}
  \psi_i(x) \to (c\tau + d)^{-k_\psi} [\rho_\psi(\gamma)]_{ij}\psi_j(x), \quad 
  \psi^c_i(x) \to (c\tau + d)^{-k_{\psi^c}} [\rho_{\psi^c}(\gamma)]_{ij}\psi^c_j(x).
\end{equation}

\noindent Fermion masses arise from Yukawa interactions of Dirac fermions. The modular-invariant Yukawa Lagrangian takes the form  
\begin{equation}
    \mathcal{L}_Y = Y^{(k_Y)}(\tau)\, \psi^c \psi H + \text{h.c.},
\end{equation}
\noindent where $H$ denotes the Higgs field (or its conjugate) and fermion fields are written in two-component notation. Under modular transformations, the Higgs field transforms as  
\begin{equation}
  H(x) \to (c\tau + d)^{-k_H} \rho_H(\gamma)\, H(x).
\end{equation}

\noindent The Yukawa couplings $Y^{(k_Y)}(\tau)$ are polyharmonic Maa\ss{} forms of weight $k_Y$ at level $N$, transforming as  
\begin{equation}
    Y^{(k_Y)}(\gamma\tau) = (c\tau + d)^{k_Y} Y^{(k_Y)}(\tau), \quad 
    \gamma = 
    \begin{pmatrix}
        a & b \\ c & d
    \end{pmatrix} \in \Gamma(N),
\end{equation}
\noindent and satisfying the Laplace equation and growth conditions:
\begin{equation}
    -4y^2 \frac{\partial}{\partial \tau} \frac{\partial}{\partial \bar{\tau}} Y^{(k_Y)} 
    + 2i k_Y y \frac{\partial}{\partial \bar{\tau}} Y^{(k_Y)} = 0, 
    \quad 
    Y^{(k_Y)}(\tau) = O(y^{\alpha}) \ \text{as } y \to +\infty.
\end{equation}

\noindent Modular invariance of the Yukawa interaction requires that the total modular weight should vanish and  the product of representations contains a singlet:
\begin{equation}
    k_Y = k_{\psi^c} + k_\psi + k_H, \quad 
    \rho_Y \otimes \rho_{\psi^c} \otimes \rho_\psi \otimes \rho_H \supset 1.
\end{equation}

\noindent For Majorana fermions $\psi^c$, the corresponding mass term is written as  
\begin{equation}
{\cal L}_M = Y^{(k_Y)}(\tau)\, \psi^c \psi^c + \text{h.c.},
\end{equation}
\noindent with modular invariance conditions  
\begin{equation}
k_Y = 2k_{\psi^c}, \quad 
\rho_Y \otimes \rho_{\psi^c} \otimes \rho_{\psi^c} \supset 1\;.
\end{equation}

\begin{table}[h!]
    \centering
    \begin{tabular}{l l}
    \hline
    Weight $k_Y$ & Polyharmonic Maa\ss{} forms $Y_r^{(k_Y)}$ \\
    \hline
    $k_Y = -2$ & $Y_1^{(-2)}$, \quad $Y_3^{(-2)}$ \\
    $k_Y = 0$  & $Y_1^{(0)}$, \quad $Y_3^{(0)}$ \\
    \hline
    \end{tabular}
    \caption{Polyharmonic Maa\ss{} forms of weight $k_Y = -2, 0$ at level $N = 3$. 
    The subscript $r$ denotes the transformation property under $A_4$ modular symmetry.}
    \label{mod_couplings}
\end{table}

\section{Model Framework}
\label{sec:linear}

This model represents the simplistic scenario of a linear seesaw, where the particle content and group charges are provided in Table 2. We prefer to extend it with a discrete $A_4$ modular non-holomorphic symmetry to explore the neutrino phenomenology. The particle spectrum is enriched with six extra singlet heavy fermion fields ($N_{Ri}$ and $S_{Li}$) and one flavon field ($\phi$). The BSM fields of the model transform as a triplet under the $A_4$ modular symmetry group. Modular symmetry is broken when the complex modulus $\tau$ acquires a VEV ⟨$\tau$⟩ in the fundamental domain, fixing numerical values of modular forms $Y^{(k)}_r (\langle\tau \rangle)$ that determine flavor structures. The extra singlet flavon field is a trivial singlet under $A_4$ symmetry.
The modular weight is assigned to all the particles and denoted as $k_I$, where the choice of modular weight is half integrals in order to avoid certain unwanted terms in the Lagrangian \cite{Zhang:2025dsa,Liu:2020msy}. The importance of $A_4$ modular symmetry is the requirement of fewer flavon fields, unlike the usual $A_4$ group, since the Yukawa couplings have the non-trivial group transformation. Assignment of group charge and modular weight to the Yukawa coupling is provided in Table \ref{tab:coupling}.

\begin{table}[h!]
\label{tab:fields-linear1}
\centering
\renewcommand{\arraystretch}{1.4}
\begin{tabular}{||c||c|c|c||c|c|c||c|c||c|c||}
\hline \hline
Fields & $L_e$ & $L_\mu$ & $L_\tau$ & $e_R$ & $\mu_R$ & $\tau_R$ & $N_R$ & $S_L$ & $H$ & $\phi$ \\
\hline \hline
$SU(2)_L$ & 2 & 2 & 2 & 1 & 1 & 1 & 1 & 1 & 2 & 1 \\
\hline
$U(1)_Y$ & $-\tfrac{1}{2}$ & $-\tfrac{1}{2}$ & $-\tfrac{1}{2}$ & $-1$ & $-1$ & $-1$ & 0 & 0 & $\tfrac{1}{2}$ & 0 \\
\hline
$A_4$ & $\mathbf{1}$ & $\mathbf{1}'$ & $\mathbf{1}''$ & $\mathbf{1}$ & $\mathbf{1}'$ & $\mathbf{1}''$ & $\mathbf{3}$ & $\mathbf{3}$ & $\mathbf{1}$ & $\mathbf{1}$ \\
\hline
$k_{I}$ & 0 & 0 & 0 & 0 & 0 & 0 & 0 & $-\tfrac{5}{2}$ & 0 & $\tfrac{1}{2}$ \\
\hline \hline
\end{tabular}
\caption{Field content and their transformation under $SU(2)_L$ $\times U(1)_Y \times A_4$, and modular weights $k_{I}$.}
\label{tab:coupling}
\end{table}


\subsection{Dirac mass term for charged leptons ($M_\ell$)}

To have a simplified structure for the charged lepton mass matrix, we consider the three generations of left-handed doublets ($L_{e}, L_{\mu}, L_{\tau} $) transform as $\bm{1},\, \bm{1'},\, \bm{1''}$ respectively under the $A_4$ symmetry. The right-handed charged leptons follow a transformation of $\bm{1}, \bm{1}^{ \prime}, \bm{1}^{\prime \prime}$ under $A_4$ symmetry respectively. All of them are assigned a modular weight of zero. The relevant Lagrangian term for charged leptons is given by
\begin{align}
 \mathcal{L}_{M_\ell}  
                   &= y_{\ell_{}}^{ee}  {L}_{e}^c H ~e_R +  y_{\ell_{}}^{\mu \mu}  {L}_{\mu}^c H~ \mu_R +  y_{\ell_{}}^{\tau \tau}  {L}_{\tau}^c H~ \tau_R.
                    \label{Eq:yuk-Mell} 
\end{align}
The charged lepton mass matrix is found to be diagonal, and the couplings can be adjusted to achieve the observed charged lepton masses. The mass matrix takes the form
\begin{align}
M_\ell = \begin{pmatrix}  y_{\ell_{}}^{ee} v_H/\sqrt{2}  &  0 &  0 \\
                                       0  &  y_{\ell_{}}^{\mu \mu} v_H/\sqrt{2}  &  0 \\
                                       0  &  0  &  y_{\ell_{}}^{\tau \tau} v_H/\sqrt{2}        \end{pmatrix}  =
                     \begin{pmatrix}  m_e  &  0 &  0 \\
                                       0  &  m_\mu  &  0 \\
                                       0  &  0  &  m_\tau      \end{pmatrix}.                 
\label{Eq:Mell} 
\end{align}
Here, $m_e$, $m_\mu$, and $m_\tau$ are the observed charged lepton masses.

\subsection{Dirac and pseudo-Dirac mass terms}
Along with the transformation of lepton doublets outlined previously, the right-handed fields $N_{Ri}$ transform as a triplet under the $A_4$ modular group. Since with these charge assignments, we cannot write the standard interaction term, we introduce the Polyharmonic Maa\ss{} form couplings to transform non-trivially under the non-holomorphic $A_4$ modular group (triplets) having modular weight of $k_I=0$, as represented in Table\,\ref{mod_couplings}. We use the modular forms of the coupling as {$\bm{Y}_3^{(0)}(\tau) = \left(Y_{3,1}^{(0)},Y_{3,2}^{(0)},Y_{3,3}^{(0)}\right)$}, expressed in Eqn.~\ref{A-1}, \ref{A-2}, and \ref{A-3} (Appendix \ref{app:A}). Therefore, the invariant Dirac Lagrangian involving the active and right-handed fields can be written as
\begin{align}
 \mathcal{L}_{D}  
                       &= \alpha_D   {L}_{e}^c \widetilde{H}~ (\bm{Y_3^{(\mathbf{0})}} N_R)_{1}   + \beta_D   {L}_{\mu}^c \widetilde{H}~ (\bm{Y_3^{(\mathbf{0})}} N_R)_{1^{\prime}}
                   + \gamma_D   {L}_{\tau}^c \widetilde{H}~ (\bm{Y_3^{(\mathbf{0})}} N_R)_{1^{\prime\prime}}.                       + {\rm h.c.}, 
                   \label{Eq:yuk-MD} 
\end{align}
Here, the subscript for the operator $\bm{Y_3^{(\mathbf{0})}} N_R$ indicates $A_4$ representation constructed by the product, $\widetilde{H}=i\sigma H^*$ and $\{\alpha_D, \beta_D, \gamma_D\}$ are free parameters. Using $v_H/\sqrt{2}$, where $v_H$ is the vacuum expectation value (VEV) of $H$, and redefining $\widetilde{x}=\frac{x}{\alpha_D}$ (i.e. $x=\beta_D,~ \gamma_D$),  the resulting Dirac mass matrix is given by
\begin{align}
M_D&=\frac{v_H}{\sqrt2}\alpha_D
\left[\begin{array}{ccc}
1 & 0 & 0 \\ 
0 & \widetilde{\beta}_D& 0 \\ 
0 & 0 & \widetilde{\gamma}_D \\ 
\end{array}\right]
\left[\begin{array}{ccc}
Y_{3,1}^{(\mathbf{0})} &Y_{3,3}^{(\mathbf{0})} &Y_{3,2}^{(\mathbf{0})} \\\\ 
Y_{3,2}^{(\mathbf{0})} &Y_{3,1}^{(\mathbf{0})} &Y_{3,3}^{(\mathbf{0})} \\\\ 
Y_{3,3}^{(\mathbf{0})} &Y_{3,2}^{(\mathbf{0})} &Y_{3,1}^{(\mathbf{0})} \\ 
\end{array}\right]_{LR}.                   
\label{Eq:MD} 
\end{align}
As we also have the extra sterile fields $S_L$, which transform analogously to $N_R$ under $A_4$ modular symmetry, the pseudo-Dirac term for the light neutrinos is allowed, and the corresponding Lagrangian is given as 
\begin{align}
 \mathcal{L}_{LS}  
                   &= \left[\alpha_{LS}   {L}_{e_L} {H}~ (\bm{Y_3^{\mathbf{(-2)}}} S^c_L)_{1}   + \beta_{LS}  {L}_{\mu_L} {H}~ (\bm{Y_3^{\mathbf{(-2)}}} S^c_L)_{1^{\prime\prime}}
                   + \gamma_{LS}   {L}_{\tau_L} {H}~ (\bm{Y_3^{\mathbf{(-2)}}} S^c_L)_{1^{\prime}}\right] \frac{\phi}{\Lambda} 
                   + {\rm h.c.}, 
                   \label{Eq:yuk-LS} 
\end{align}
where, the choice of the Yukawa coupling is $\mathbf{Y}_3^{(\mathbf{-2})}=\left(Y_{3,1}^{(\mathbf{-2})}, Y_{3,2}^{(\mathbf{-2})}, Y_{3,3}^{(\mathbf{-2})}\right)$ (expressions can be found in Appendix \ref{app:A} Eqn. \ref{A-4}, \ref{A-5}, \ref{A-6}) to keep the Lagrangian invariant with the subscript for the operator $(\bm{Y_3^{\mathbf{(-2)}}}{S^c_L})$ indicates $A_4$ representation constructed by its product rule and $\{\alpha_{LS}, \beta_{LS}, \gamma_{LS}\}$ are free parameters. Defining $\widetilde{x}=\frac{x}{\alpha_{LS}}$ (for $x=\beta_{LS},\gamma_{LS}$), the flavor structure for the pseudo-Dirac neutrino mass matrix takes the form,

\begin{align}
M_{LS}&=\frac{v_H}{\sqrt2}\left(\frac{v_\phi}{\sqrt{2}\Lambda}\right) \alpha_{LS}
\left[\begin{array}{ccc}
1 & 0 & 0 \\ 
0 & \widetilde{\beta}_{LS} & 0 \\ 
0 & 0 & \widetilde{\gamma}_{LS} \\ 
\end{array}\right]
\left[\begin{array}{ccc}
Y_{3,1}^{(\mathbf{-2})} &Y_{3,3}^{(\mathbf{-2})} &Y_{3,2}^{(\mathbf{-2})} \\\\ 
Y_{3,3}^{(\mathbf{-2})} &Y_{3,2}^{(\mathbf{-2})} &Y_{3,1}^{(\mathbf{-2})} \\\\
Y_{3,2}^{(\mathbf{-2})} &Y_{3,1}^{(\mathbf{-2})} &Y_{3,3}^{(\mathbf{-2})} \\
\end{array}\right]_{LR} ,              
\label{Eq:MLS} 
\end{align}\\
where $v_\phi$ is the VEV of $\phi$, i.e.,  $\langle \phi \rangle = v_\phi/\sqrt{2}$.

\subsection{Mixing between the heavy fields $N_R$ and $S^c_L$}
Following the transformation of the heavy fields under the imposed symmetries, one can have the interactions leading to the mixing between these additional fields as follows: 
\begin{eqnarray}
 \mathcal{L}_{M_{RS}}  
                   &=& [\alpha_{NS} \mathbf{Y}_3^{(-\mathbf{2})} ({S^c_L} N_R)_{\rm sym} + \beta_{NS} \mathbf{Y}_3^{(-\mathbf{2})}({S^c_L} N_R)_{\rm Anti-sym} ]\phi  
                   + {\rm h.c.}            
     \label{Eq:yuk-M} 
\end{eqnarray}
where the first and second terms in  (\ref{Eq:yuk-M}) correspond to symmetric and anti-symmetric product for $S^c_L N_R$, making triplet representation of $A_4$ with $\alpha_{NS}$, $\beta_{NS}$ being the free parameters with $\widetilde{\beta}_{NS} = \beta_{NS}/\alpha_{NS}$.
Thus,  the resulting mass matrix is found to be,

\begin{align}
M_{RS}&=\frac{v_\phi}{\sqrt2}\alpha_{NS}
 \left(
 \frac{1}{3}\left[\begin{array}{ccc}
2 Y_{3,1}^{(\mathbf{-2})} ~&~ - Y_{3,3}^{(\mathbf{-2})} ~&~ -Y_{3,2}^{(\mathbf{-2})} \\ \\
-Y_{3,3}^{(\mathbf{-2})} ~&~ 2 Y_{3,2}^{(\mathbf{-2})} ~&~ -Y_{3,1}^{(\mathbf{-2})} \\ \\
-Y_{3,2}^{(\mathbf{-2})} ~&~ -Y_{3,1}^{(\mathbf{-2})} ~&~ 2 Y_{3,3}^{(\mathbf{-2})} \\ 
\end{array}\right]
+
\widetilde{\beta}_{NS}
\left[\begin{array}{ccc}
0 ~&~Y_{3,3}^{(\mathbf{-2})} ~&~ -Y_{3,2}^{(\mathbf{-2})} \\ \\
-Y_{3,3}^{(\mathbf{-2})}~&~ 0 ~&~ Y_{3,1}^{(\mathbf{-2})} \\ \\
Y_{3,2}^{(\mathbf{-2})} ~&~ -Y_{3,1}^{(\mathbf{-2})} ~&~0 \\ 
\end{array}\right]
\right). \label{yuk:MRS}
\end{align}
It should be noted that $1/3 \neq \widetilde{\beta}_{NS}$, otherwise the matrix $M_{RS}$  becomes singular, which eventually spoils the intent of the linear seesaw. 
\subsection{Majorana term for $N_R$}
The Majorana mass term for $N_R$ is allowed in this framework, which can be obtained from the Lagrangian.

\begin{equation}
    \mathcal{L}_N = \frac{M_{R}}{2} \left( \alpha_R (N_R \overline{N^c_R}) {\mathbf Y_3^{(\mathbf{0})}}+  \beta_R (N_R \overline{N^c_R}) {\mathbf Y_1^{(\mathbf{0})}} \right) + h.c.
\end{equation}
and is expressed as
\begin{equation}
   M_N = M_R \alpha_R 
   \begin{pmatrix}
       2 Y_{3,1}^{(\mathbf{0})}+ \widetilde{\beta}_R ~&~ -Y_{3,3}^{(\mathbf{0})} ~&~ -Y_{3,2}^{(\mathbf{0})}\\\\
       -Y_{3,3}^{(\mathbf{0})} ~&~ 2Y_{3,2}^{(\mathbf{0})} ~&~ -Y_{3,1}^{(\mathbf{0})}+\widetilde{\beta}_R \\\\
       -Y_{3,2}^{(\mathbf{0})} ~&~ -Y_{3,1}^{(\mathbf{0})}+\widetilde{\beta}_R  ~&~ 2Y_{3,3}^{(\mathbf{0})}
   \end{pmatrix}.
\end{equation}
with $\mathbf {Y_1^{(0)}}=1$ and $\widetilde{\beta}_R = \beta_R/\alpha_R$.
\subsection{Linear Seesaw mechanism for light neutrino masses}
Within the present model invoked with $A_4$ modular symmetry, the complete $9 \times 9$ mass matrix  in the flavor basis of $\left(\nu_L, N_R, S^c_L \right)^T$ is given by
\begin{eqnarray}
\mathbb{M} = \left(\begin{array}{c|ccc}   & \nu_L & N_R  & S^c_L   \\ \hline
\nu_L  & 0       & M_D       & M_{LS} \\
N_R    & M^T_D         & M_N       & M_{RS} \\
S^c_L & M_{LS}^T     & M_{RS}^T    & 0
\end{array}
\right),
\label{eq:numatrix-complete}
\end{eqnarray}

The effective light neutrino mass matrix can be obtained by block 
diagonalizing Eq.~(\ref{eq:numatrix-complete}). 
In this linear seesaw framework{\footnote{The flavon $\phi$ with fractional charge allows the retention of the mixing between $N_R$ and $S_L$, enabling the form of Eq.~(3.11). }} the expansion is performed under 
the hierarchical regime
\begin{equation}
\label{eq:hierarchy}
M_{LS} \ll M_D \ll M_{RS}, 
\qquad 
M_N \ll M_{RS}, 
\end{equation}
where the heavy $(N_R, S_L^c)$ block is dominated by $M_{RS}$ and the lepton number violating terms $M_{LS}$ and $M_N$ are expected to be naturally small. 
This ensures that the eigenvalues of the heavy sector are of 
$\mathcal{O}(M_{RS})$, justifying the seesaw expansion.

Block diagonalization of Eq.~(\ref{eq:numatrix-complete}) (see Appendix \ref{appex-B}), yields the effective light 
neutrino mass matrix, having the form
\begin{eqnarray}
m_\nu 
&=& M_D (M_{LS}M_{RS}^{-1})^T 
+ (M_{LS}M_{RS}^{-1}) M_D^T  
 -\, M_{LS} M_{RS}^{-1} M_N (M_{RS}^T)^{-1} M_{LS}^T \;.
\label{eq:mnu-full}
\end{eqnarray}
Under the hierarchical conditions mentioned in Eq. (\ref{eq:hierarchy}), the last term in (\ref{eq:mnu-full}) turns out to be of the form of double seesaw and is considerably suppressed with respect to the first two terms and acts as a higher-order correction, which can be safely ignored. 

Thus, retaining only the leading contribution, the light neutrino mass matrix takes the linear seesaw form
\begin{equation}
m_\nu \simeq M_D M_{RS}^{-1} M_{LS}^T + \text{transpose}
= \left[\frac{v_H^2 \alpha_D \alpha_{LS}}{2\Lambda \hspace*{0.05 truecm} \alpha_{NS}}\right] \widetilde{m}_\nu
= \kappa\, \widetilde{m}_\nu,
\end{equation}
where $\kappa = \frac{v_H^2 \alpha_D \alpha_{LS}}{2\Lambda \hspace*{0.05 true cm}\alpha_{NS}}\ $ sets the overall mass scale and $\widetilde{m}_\nu$ is a dimensionless $3\times 3$ matrix. The Hermitian matrix $\widetilde{m}_\nu^\dagger \widetilde{m}_\nu$ is diagonalized by a unitary matrix $V_\nu$ as
\begin{equation}
V_\nu^\dagger \left( \widetilde{m}_\nu^\dagger \widetilde{m}_\nu \right) V_\nu
= \text{diag}\left( \widetilde{D}_{\nu_1}^2,\widetilde{D}_{\nu_2}^2,\widetilde{D}_{\nu_3}^2 \right),
\qquad
V_\nu^\dagger V_\nu = 1_{3\times3}.
\end{equation}
The overall scale $\kappa$ is fixed by the atmospheric mass-squared difference:
\begin{equation}
(\text{NO}): \quad
\kappa^2 =
\frac{|\Delta m^2_{\text{atm}}|}
{\widetilde{D}_{\nu_3}^2 - \widetilde{D}_{\nu_1}^2},
\qquad
(\text{IO}): \quad
\kappa^2 =
\frac{|\Delta m^2_{\text{atm}}|}
{\widetilde{D}_{\nu_2}^2 - \widetilde{D}_{\nu_3}^2},
\end{equation}
where NO (IO) denotes normal (inverted) ordering. The solar mass-squared difference is then given by
\begin{equation}
\Delta m^2_{\text{sol}}
= \kappa^2
\left( \widetilde{D}_{\nu_2}^2 - \widetilde{D}_{\nu_1}^2 \right),
\end{equation}
which can be compared with the observed value. 
\subsection{Neutrino Oscillation Observables}
\label{Num_ana}
Neutrino oscillations are described by six physical observables: the two
mass-squared differences $\Delta m^2_{\rm sol}$ and $\Delta m^2_{\rm atm}$,
three mixing angles $\theta_{12}$, $\theta_{23}$, $\theta_{13}$, and the Dirac
CP phase $\delta_{\rm CP}$. The mass-squared splittings are defined as
$\Delta m^2_{\rm sol} = m_2^2 - m_1^2$ for both normal ordering (NO) and
inverted ordering (IO), while
$\Delta m^2_{\rm atm} = m_3^2 - m_1^2$ for NO and
$\Delta m^2_{\rm atm} = m_2^2 - m_3^2$ for IO.
The leptonic mixing angles are extracted from the elements of the PMNS matrix $U$ as
\begin{align}
\sin^2\theta_{12} &= \frac{|U_{e2}|^2}{1 - |U_{e3}|^2}, \\
\sin^2\theta_{23} &= \frac{|U_{\mu 3}|^2}{1 - |U_{e3}|^2}, \\
\sin^2\theta_{13} &= |U_{e3}|^2 ,
\end{align}
while the Dirac CP phase can be determined through the rephasing-invariant product 
$U_{e2} U_{\mu3} U_{e3}^\ast U_{\mu2}^\ast$ 
of the PMNS matrix elements.

The allowed parameter space is obtained by fitting these inputs to the six neutrino oscillation observables using results from global analyses~\cite{deSalas:2020pgw,Capozzi:2021fjo,Esteban:2024eli}. The
corresponding best-fit values and $3\sigma$ ranges are summarized in
Table~\ref{table:data_nufit}. Apart from the smallness of neutrino masses, another important observable in the neutrino sector includes the Jarlskog invariant, which quantifies CP violation in neutrino oscillations and can be expressed in terms of the mixing angles and CP-violating phases of the PMNS matrix as 
\begin{eqnarray}
 J_{CP} = \text{Im} [U_{e1} U_{\mu 2} U_{e 2}^* U_{\mu 1}^*] = \sin\theta_{23} \cos\theta_{23} \sin\theta_{12} \cos\theta_{12} \sin\theta_{13} \cos^2\theta_{13} \sin \delta_{CP}.~
\end{eqnarray}

\begin{table}[ht]
    \centering
    \begin{tabular}{|l|c|c|c|c|c|l|}
    \hline\hline
\multirow{2}{*}{Parameters}  & \multicolumn{2}{c|}{Normal ordering}  & \multicolumn{2}{c|}{Inverted ordering} \\\cline{2-5}
    &Best-fit value & $3\sigma$ range &  Best-fit value & $3\sigma$ range \\ \hline
$\sin^2\theta_{12}$ & $0.308_{-0.011}^{+0.012}$  & $0.275 - 0.345$ & $0.308_{-0.011}^{+0.012}$ & $0.275 - 0.345$ \\
$\sin^2\theta_{23}$ & $0.470_{-0.013}^{+0.017}$ & $0.435 - 0.585$ & $0.550_{-0.015}^{+0.012}$ & $0.440 - 0.584$\\
$\sin^2\theta_{13}$ & $0.02215_{-0.00058}^{+0.00056}$ & $0.02030 - 0.02388$ & $0.02231_{-0.00056}^{+0.00056}$ & $0.02060 - 0.02409$ \\
\hline 
$\frac{\Delta m_{\mathrm{sol}}^2}{10^{-5} \mathrm{eV^2}}$ & $7.49_{-0.19}^{+0.19}$ & $6.92- 8.05$ & $7.49_{-0.19}^{+0.19}$ & $6.92- 8.05$\\
$\frac{\Delta m_{\mathrm{atm}}^2}{10^{-3} \mathrm{eV^2}}$
        & $2.513_{-0.019}^{+0.021}$ & $2.451- 2.578$ & $2.484_{-0.020}^{+0.020}$ & $ 2.421-2.547$ \\
        \hline
        $\delta_{\mathrm{CP}}/^{\circ}$ & $212_{-41}^{+26}$ & $ 124- 364$  & $ 274_{-25}^{+22}$  & $201 - 335$ \\
\hline
\hline
\end{tabular}
    \caption{Best-fit values and their $3\sigma$ ranges for neutrino oscillation parameters obtained from NuFIT 6.0~\cite{Esteban:2024eli}.}
    \label{table:data_nufit}
\end{table}

\section{Simulation Details}
\label{simulation}

The  simulations for the long-baseline neutrino experiments are performed using the GLoBES framework~\cite{Huber:2004ka,Huber:2007ji}. For the DUNE setup, we employ the official configuration files corresponding to the Technical Design Report (TDR)~\cite{DUNE:2021cuw}. DUNE is a next-generation accelerator-based neutrino oscillation experiment based at Fermilab (USA), with its far detector located in South Dakota. The experiment utilizes a high-intensity neutrino beam with a nominal beam power of 1.2~MW and covers a broad neutrino energy range. The far detector consists of a 40~kt liquid argon time projection chamber (LArTPC), while the near detector complex is situated at Fermilab to characterize the beam.

In our analysis, we assume a total exposure of 13 years, divided equally between neutrino and antineutrino modes (6.5 years each). This corresponds to an annual exposure of $1.1 \times 10^{21}$ protons on target (PoT). The adopted configuration is consistent with the nominal staging scenario described in Ref.~\cite{DUNE:2020jqi}.

The experimental sensitivity is evaluated through a $\chi^2$ analysis based on the Poisson log-likelihood function,
\begin{equation}
\chi^2_{\rm stat} = 2 \sum_{i=1}^{n}
\left[
N^{\rm test}_i - N^{\rm true}_i
- N^{\rm true}_i \ln\left(
\frac{N^{\rm test}_i}{N^{\rm true}_i}
\right)
\right],
\end{equation}
where $N^{\rm true}_i$ and $N^{\rm test}_i$ denote the predicted event rates in the $i^{\rm th}$ energy bin for the true and test hypotheses, respectively, and $n$ represents the total number of bins.

Systematic uncertainties are incorporated using the pull method~\cite{Fogli:2002pt,Huber:2002mx}. In our implementation, both signal and background normalization uncertainties are included, along with shape uncertainties where applicable. The systematic inputs adopted for DUNE are summarized in Table~\ref{table_sys}.

The oscillation parameters are taken from the latest global fit results of NuFit~6.1~\cite{Esteban:2024eli}, listed in Table~\ref{table:data_nufit}. During the statistical analysis, we marginalize over the oscillation parameters within the ranges predicted by the underlying model.

To assess the compatibility of the model predictions with the NuFit~6.1 results, 
we compute an additional $\chi^2$ defined as
\begin{equation}
\chi^2 = \sum_i 
\frac{\left(O^{\rm Th}_i - O^{\rm Exp}_i\right)^2}
{\sigma_i^2},
\label{chi-th}
\end{equation}
where $O^{\rm Th}_i$ and $O^{\rm Exp}_i$ denote the theoretical and experimental central values of the $i^{\rm th}$ observable, respectively, and $\sigma_i$ corresponds to the associated experimental uncertainty.

\begin{table} 
\centering
\begin{tabular}{|c|c|c|} \hline
Systematics       & DUNE  \\ \hline
Sg-norm $\nu_{e}$     & 2$\%$      \\ 
Sg-norm $\nu_{\mu}$             & 5$\%$ \\ 
Bg-norm      & 5$\%$ to 20$\%$\\ 
\hline
\end{tabular}
\caption{Systematic uncertainties considered in the DUNE simulation. Here ``Sg'' and ``Bg'' denote signal and background contributions, respectively, while ``norm'' refers to normalization uncertainties.}
\label{table_sys}
\end{table}

\section{Results}
\label{sec:results}
In this section, we identify the allowed region of our model parameters consistent with neutrino oscillation data at the 3$\sigma$ level, see Tab.~\ref{table:data_nufit}. Recall that in our setup, we have five free parameters along with the modulus $\tau$. In our scan, $\tau$ is varied within the fundamental domain defined by 
\begin{equation}
    \Big|\text{Re}(\tau)\Big|\in [0,0.5],\quad {\rm Im}(\tau)\in [0.86,5.0], 
\end{equation} 
while the ratio of the free parameters are taken to be in the range
\begin{equation}
    \tilde\beta_{D},\tilde\beta_{LS},\tilde\gamma_{D},\tilde\gamma_{LS},\tilde\beta_{NS} \in \left[10^{-3}, 10^3\right].
\end{equation}
For fitting the neutrino oscillation observables and we make use of a package Flavorpy \cite{FlavorPy}.

\subsection{Predictions from the model}
\begin{figure}
\centering

\includegraphics[width=1\linewidth]{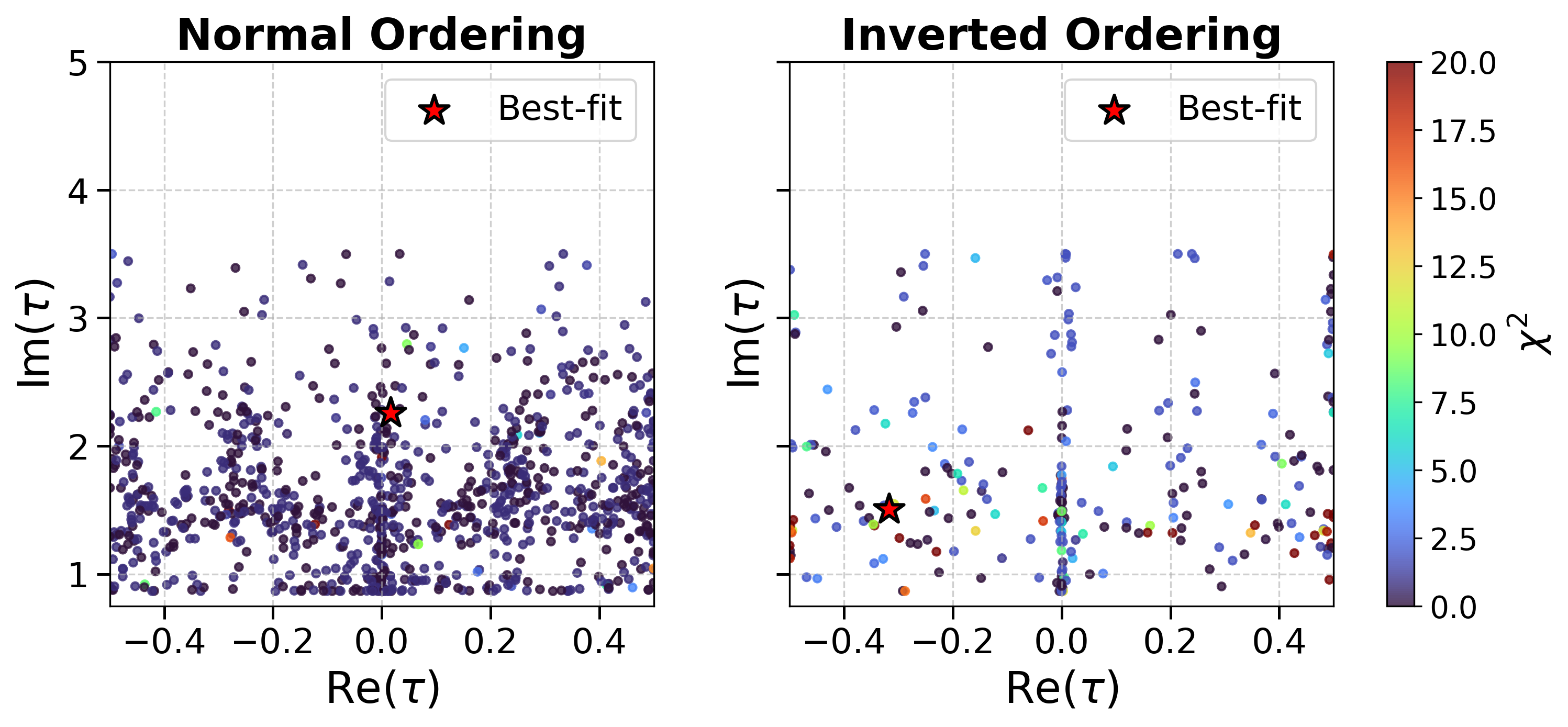}
\caption{Parameter space between Re($\tau$) and Im($\tau$) for the model preferring NO and IO in left and right panel, respectively. The colour regions represent the different $\chi^2$ values.}
\label{fig:ret-imt}
\end{figure}

Our model accommodates both Normal Ordering (NO) as well as Inverted Ordering (IO) of neutrino masses for phenomenologically viable values of the complex modulus parameter $\tau$. The allowed regions in the complex $\tau$-plane are shown in Fig.~\ref{fig:ret-imt}, where we scan the parameter space by varying $\mathrm{Re}(\tau)$ in the range $[-0.5,\,0.5]$ and $\mathrm{Im}(\tau)$ in the range $[0.86,\,5]$. Best fit points are shown with  red asterisk symbol, and the corresponding best-fit values of $\mathrm{Re}(\tau)$ and $\mathrm{Im}(\tau)$ are summarized in Table~\ref{tab:best-fit-param}. For the normal ordering case, the analysis yields best-fit values $\mathrm{Re}(\tau)=0.02$ and $\mathrm{Im}(\tau)=2.26$. In contrast, for the inverted ordering scenario, the model prefers comparatively smaller values of the modulus parameter, with $\mathrm{Re}(\tau)=-0.32$ and $\mathrm{Im}(\tau)=1.5$. 

\begin{table}[h!]
\centering
\renewcommand{\arraystretch}{1.1}
\setlength{\tabcolsep}{4pt}
\small
\begin{tabular}{|c|c|c|c|c|}
\hline
\multirow{2}{*}{\textbf{Parameters}} 
 & \multicolumn{2}{c|}{\textbf{Normal Ordering (NO)}} 
 & \multicolumn{2}{c|}{\textbf{Inverted Ordering (IO)}} \\ \cline{2-5}
 & Best fit  & Parameter range & Best fit  & Parameter range \\ \hline
 Re($\tau$) &$0.02$ & [-0.5, 0.5]& $-0.32$ & [-0.5,0.5] \\ \hline
 Im ($\tau$) &$2.26$& [0.86, 5]&   $1.5$ & [0.86,5] \\ \hline
\end{tabular}
\caption{Best fit values of the  parameter from the model for NO and IO.}
\label{tab:best-fit-param}
\end{table}

\begin{figure}[htbp]
\centering
\includegraphics[width=.45\textwidth]{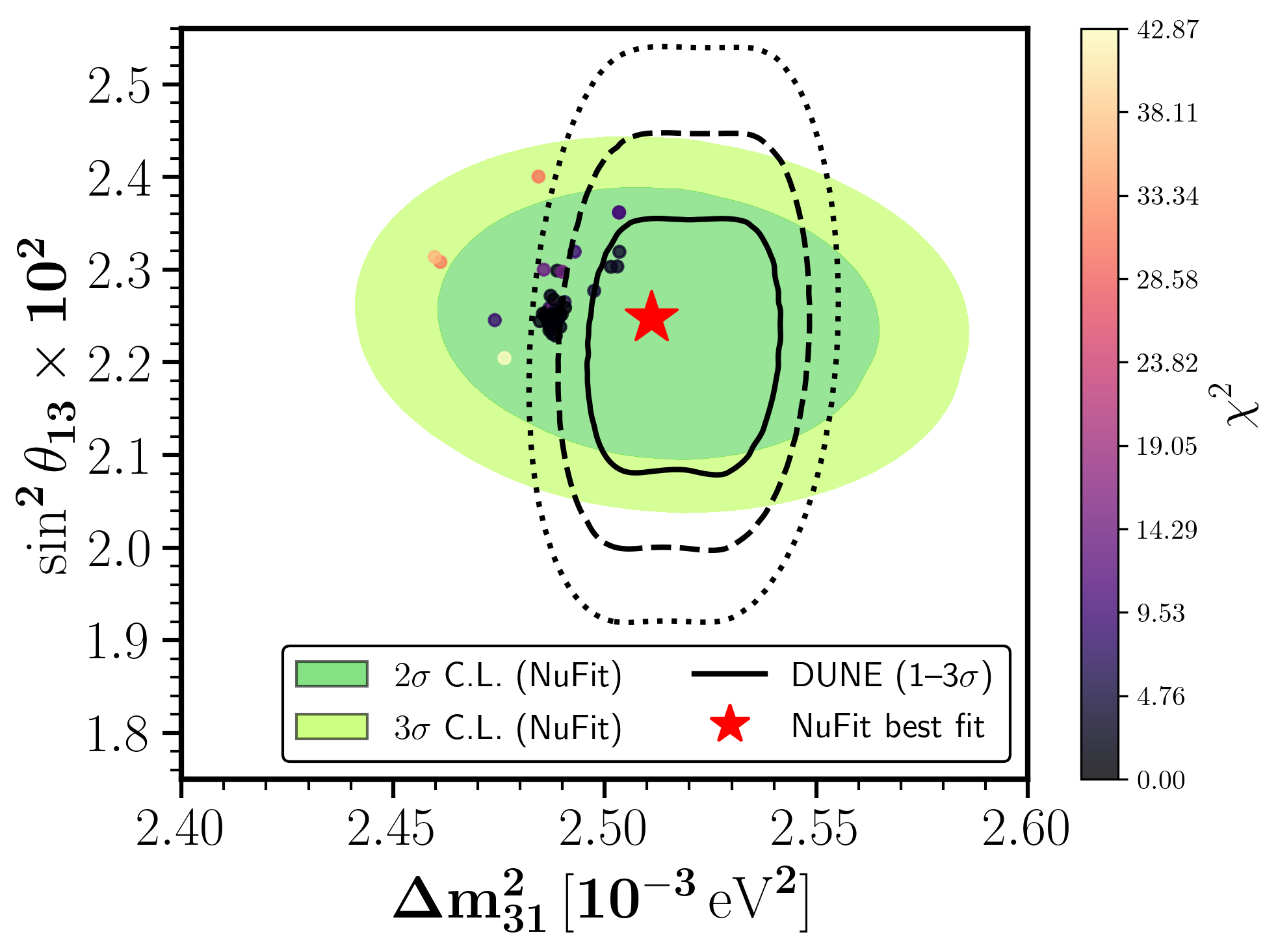}
\includegraphics[width=.45\textwidth]{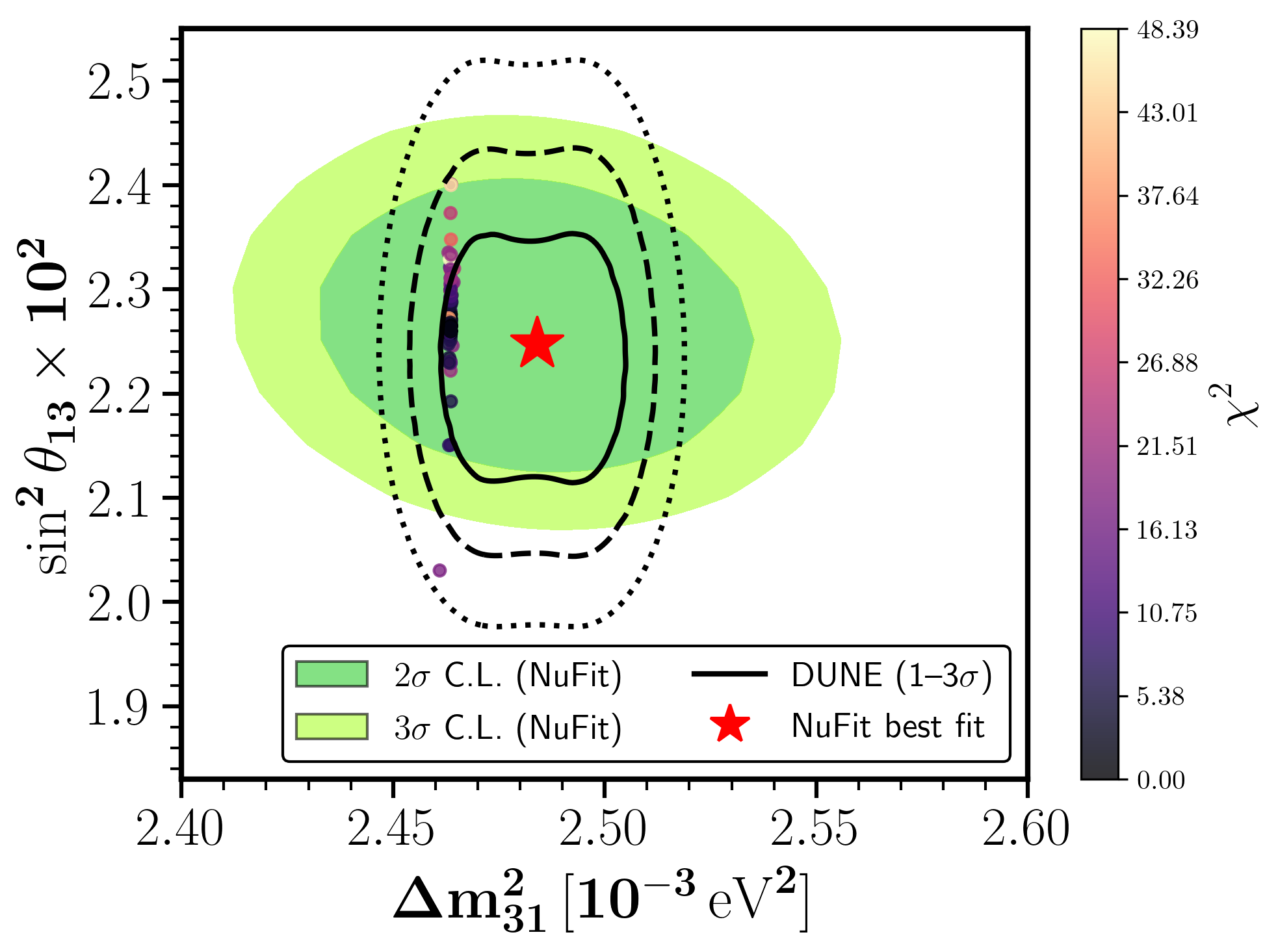}\\
\includegraphics[width=.45\textwidth]{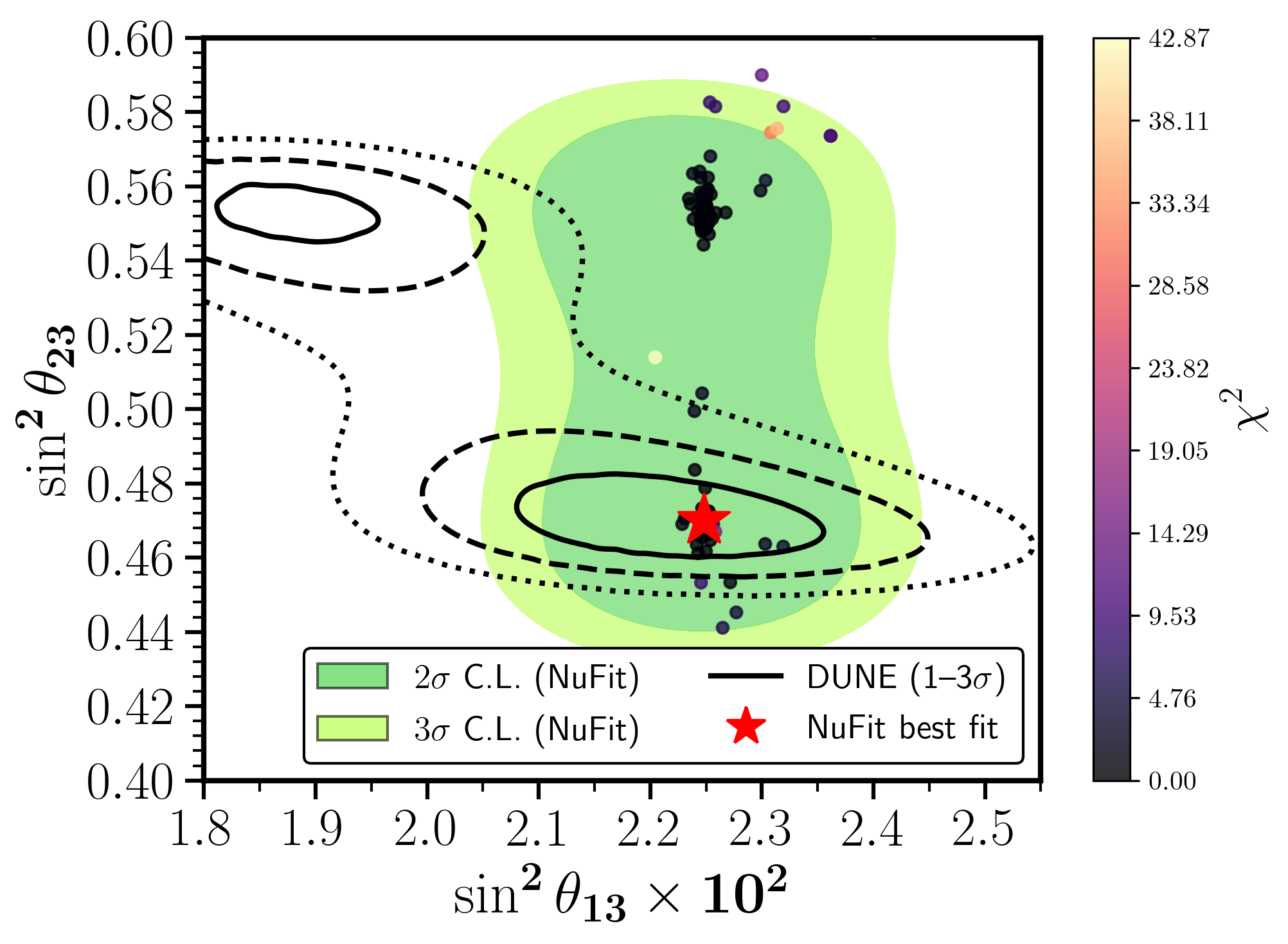}
\includegraphics[width=.45\textwidth]{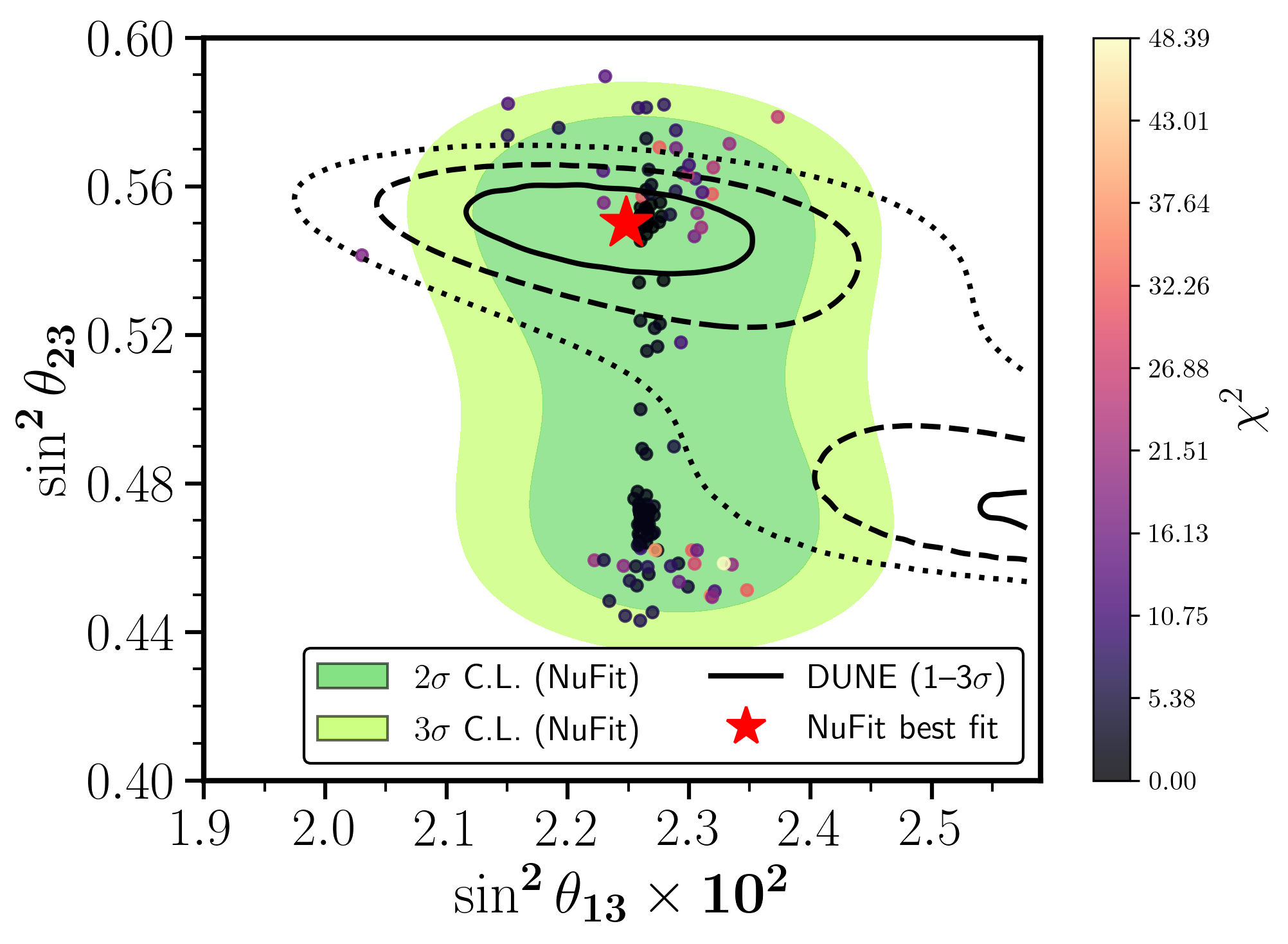}\\
\includegraphics[width=.45\textwidth]{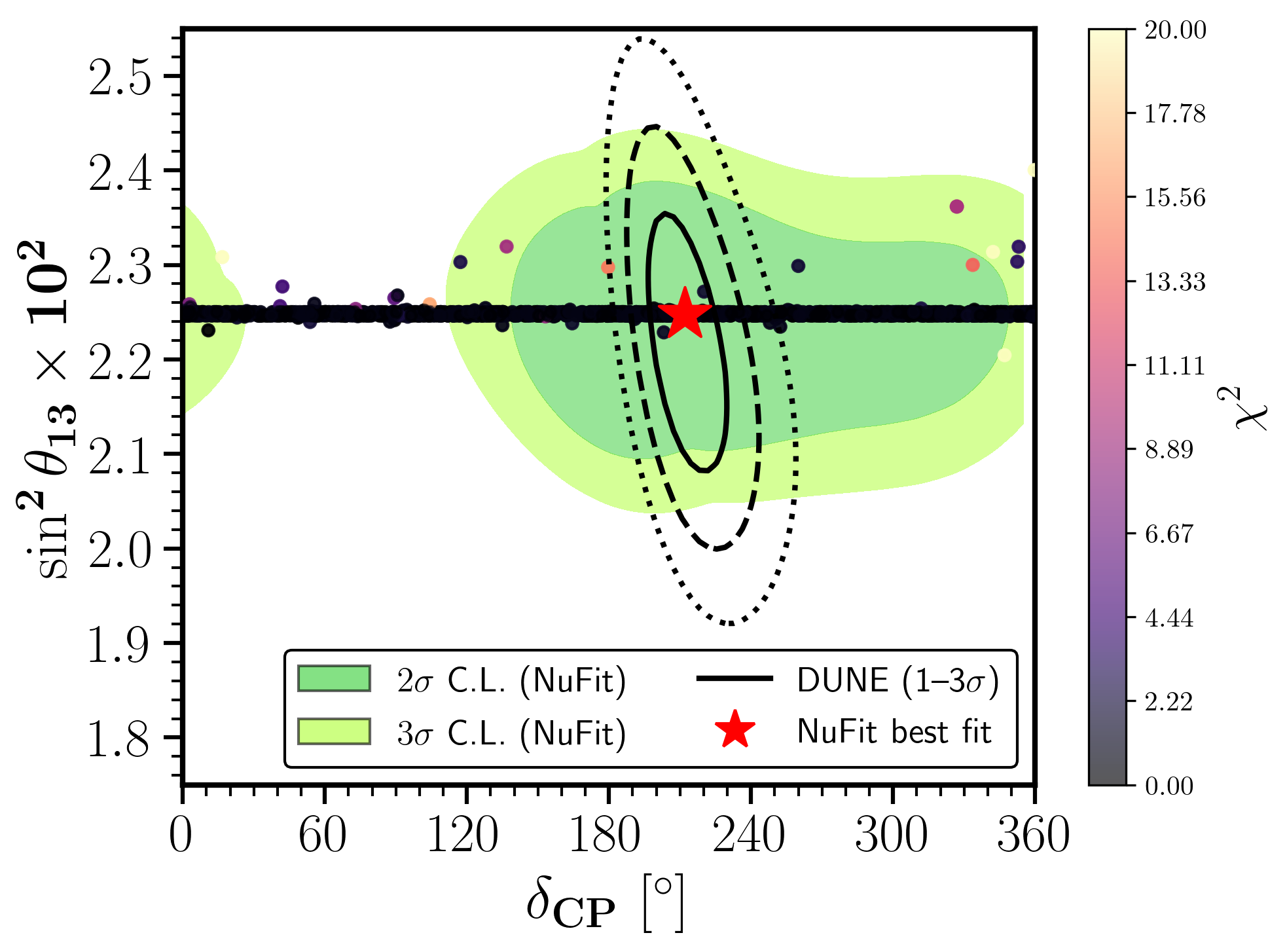}
\includegraphics[width=.45\textwidth]{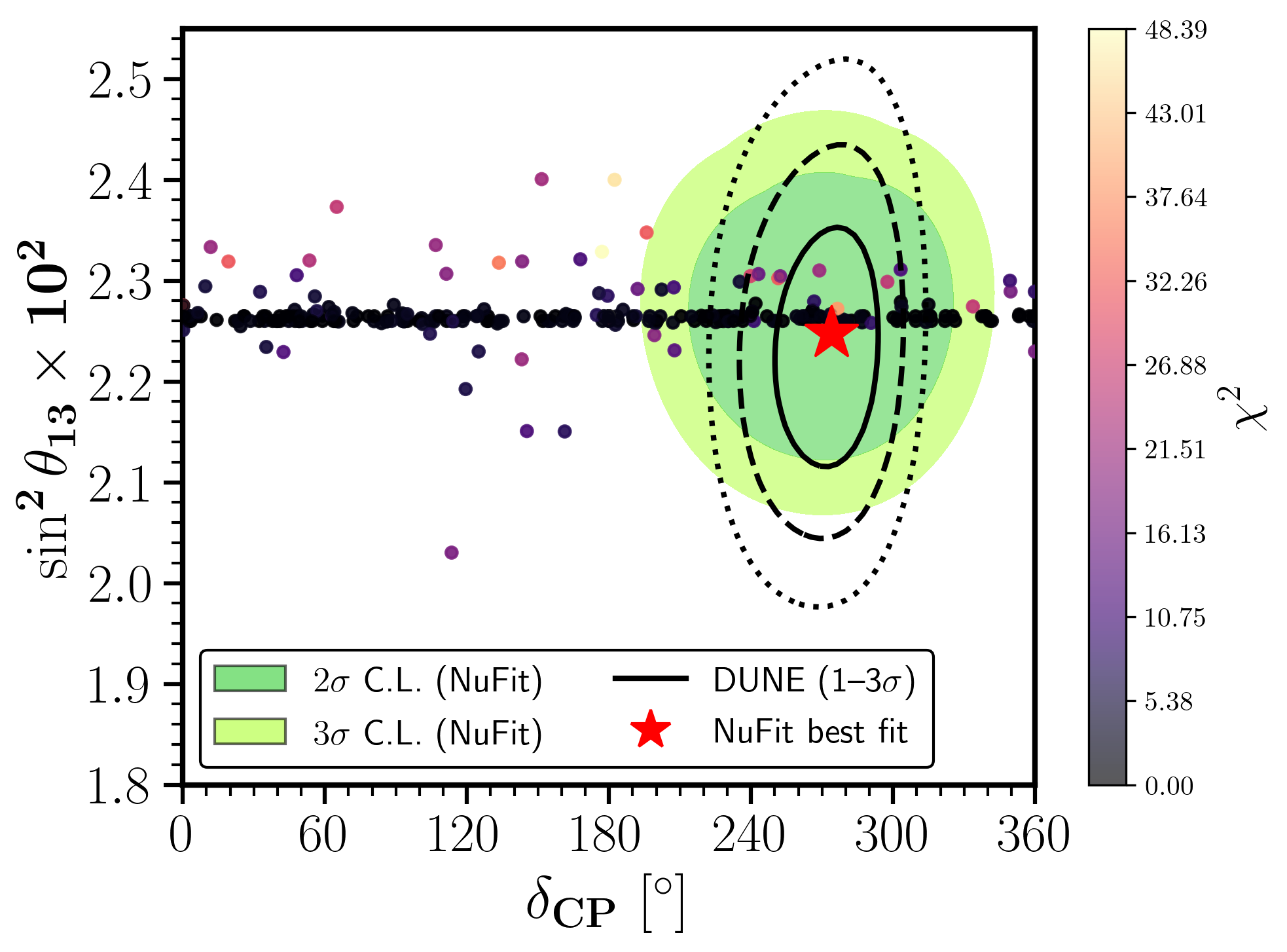}
\caption{Allowed parameter spaces among the oscillation parameters for normal ordering (NO, left panel) and inverted ordering (IO, right panel). The contour lines indicate the allowed regions from the DUNE experiment, while the green contours show the NuFIT constraints. The scattered points correspond to the parameter space permitted by the model.}
\label{fig:allowed-no-io}
\end{figure}

\begin{figure}[htbp]
\centering
 \includegraphics[width=.45\textwidth]{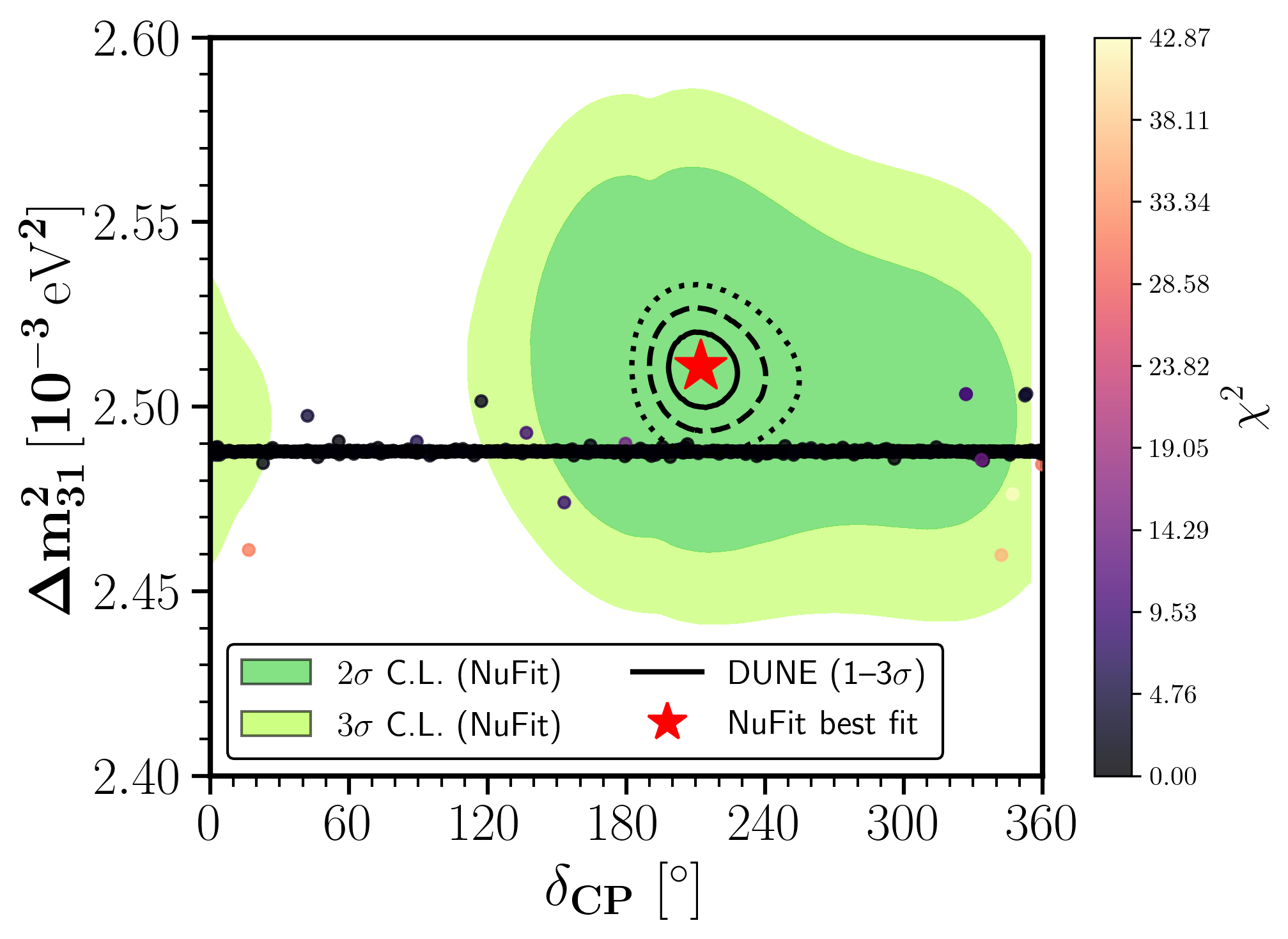}
\includegraphics[width=.45\textwidth]{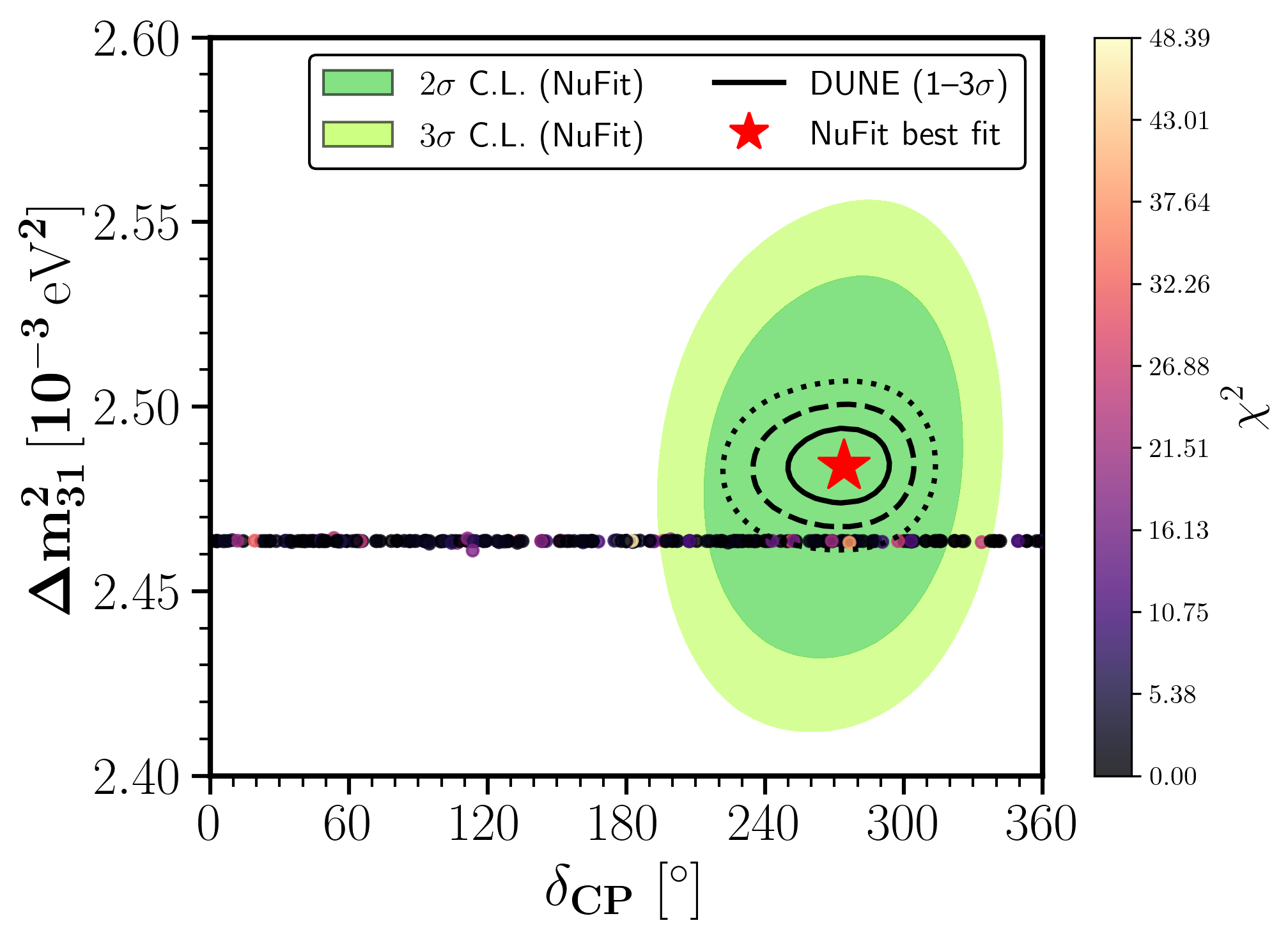}\\
\includegraphics[width=.45\textwidth]{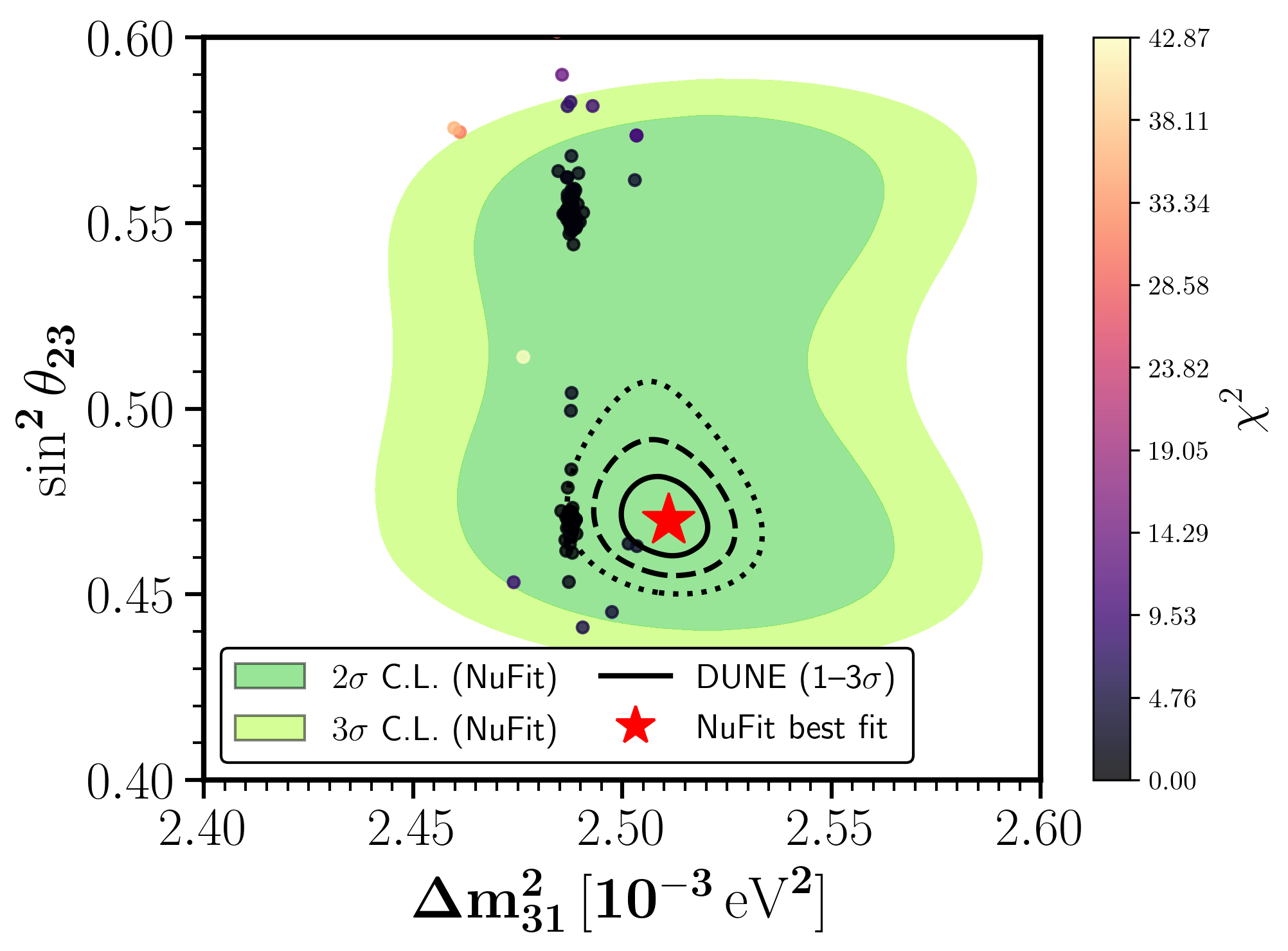}
\includegraphics[width=.45\textwidth]{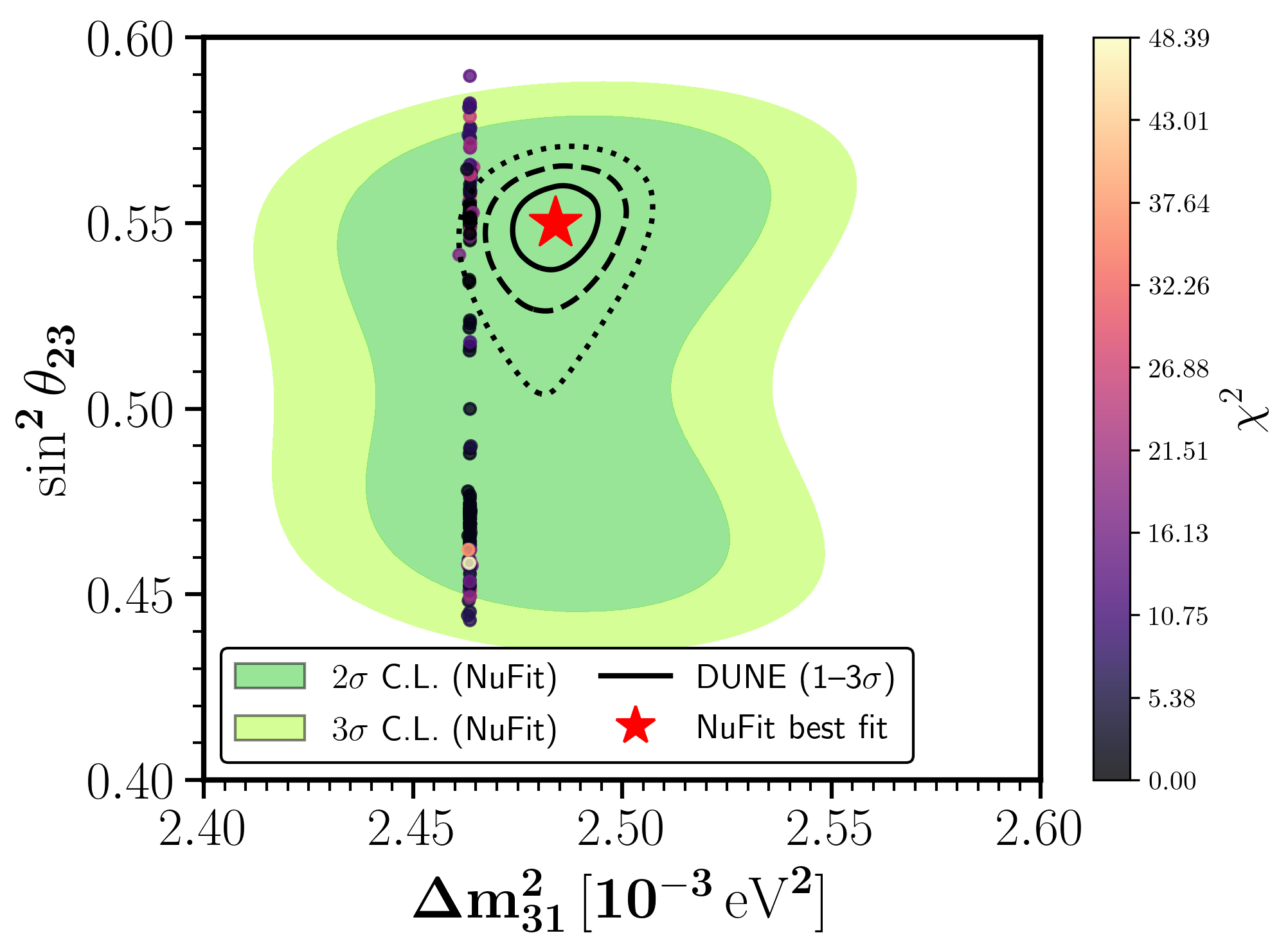}\\
      \includegraphics[width=0.48\linewidth]{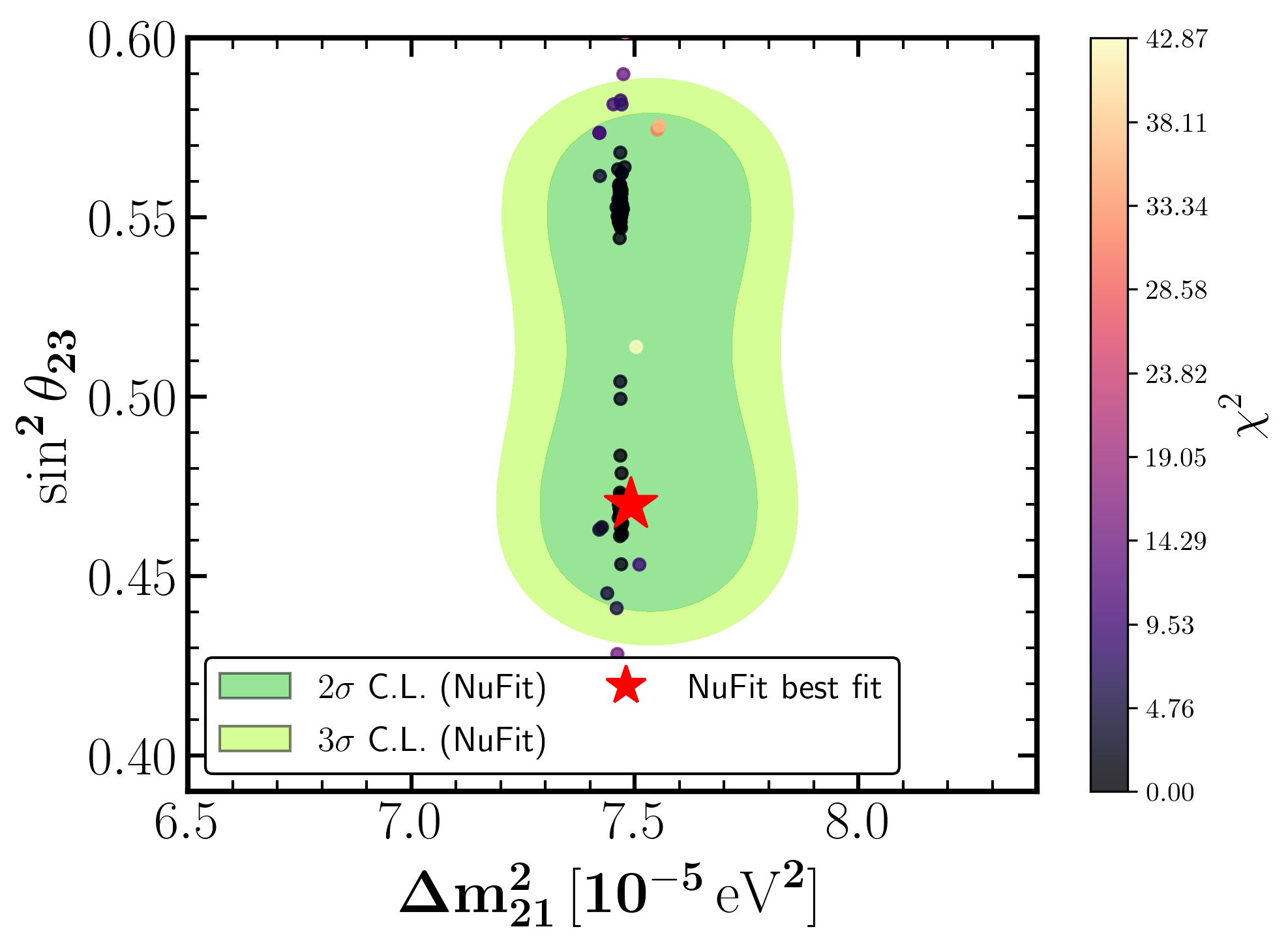}
    \includegraphics[width=0.48\linewidth]{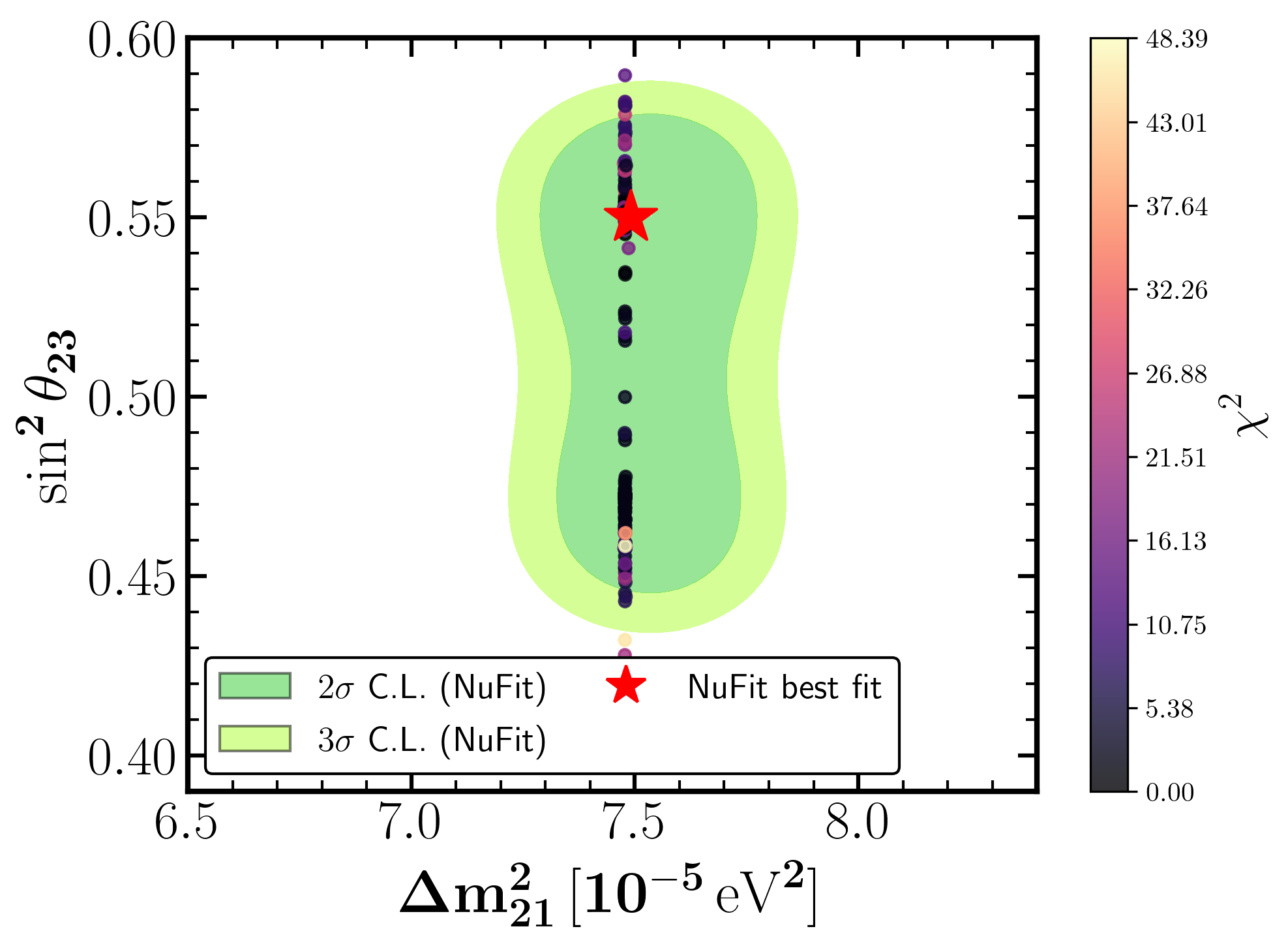}
\includegraphics[width=.45\textwidth]{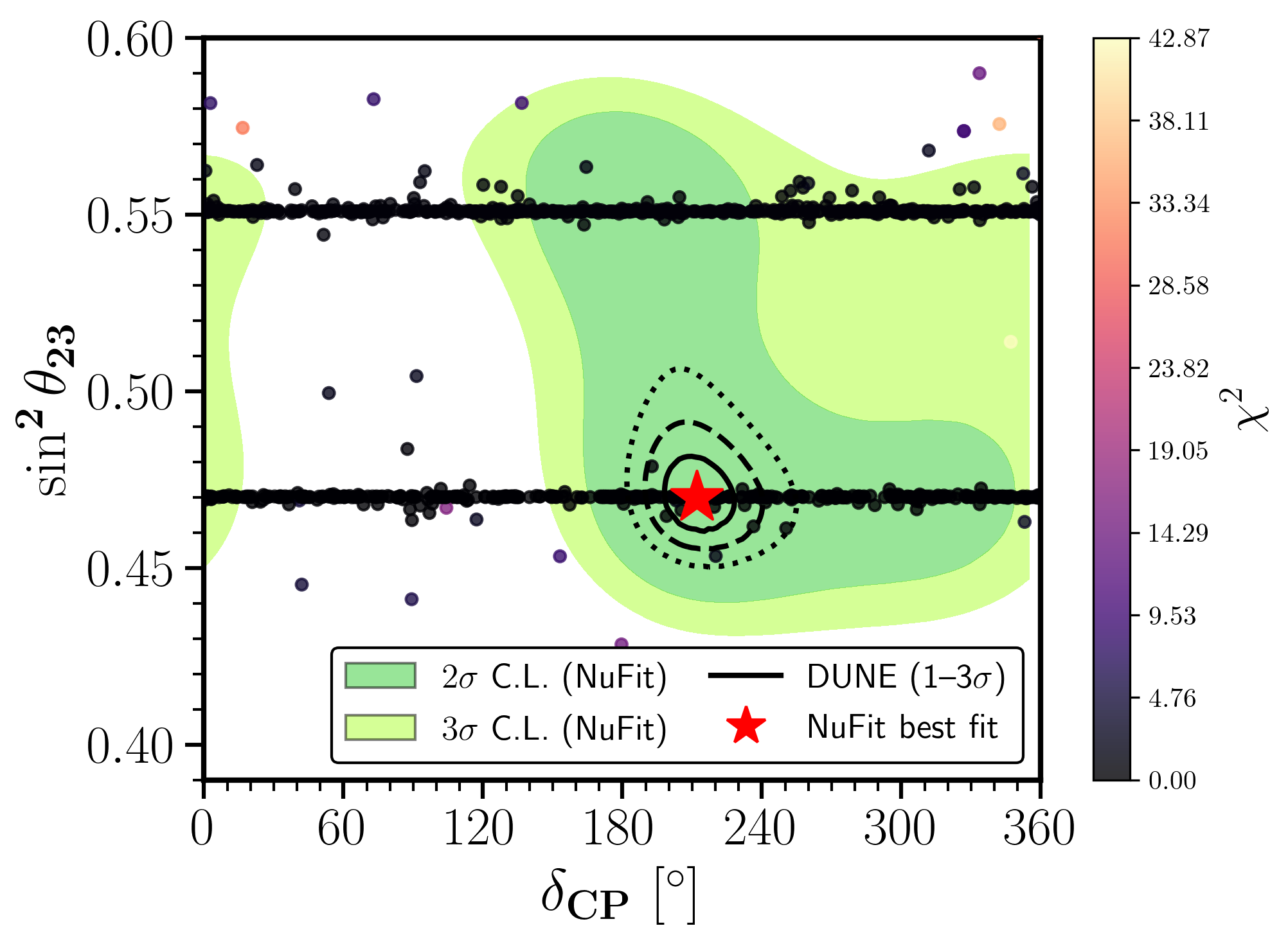}
\includegraphics[width=.45\textwidth]{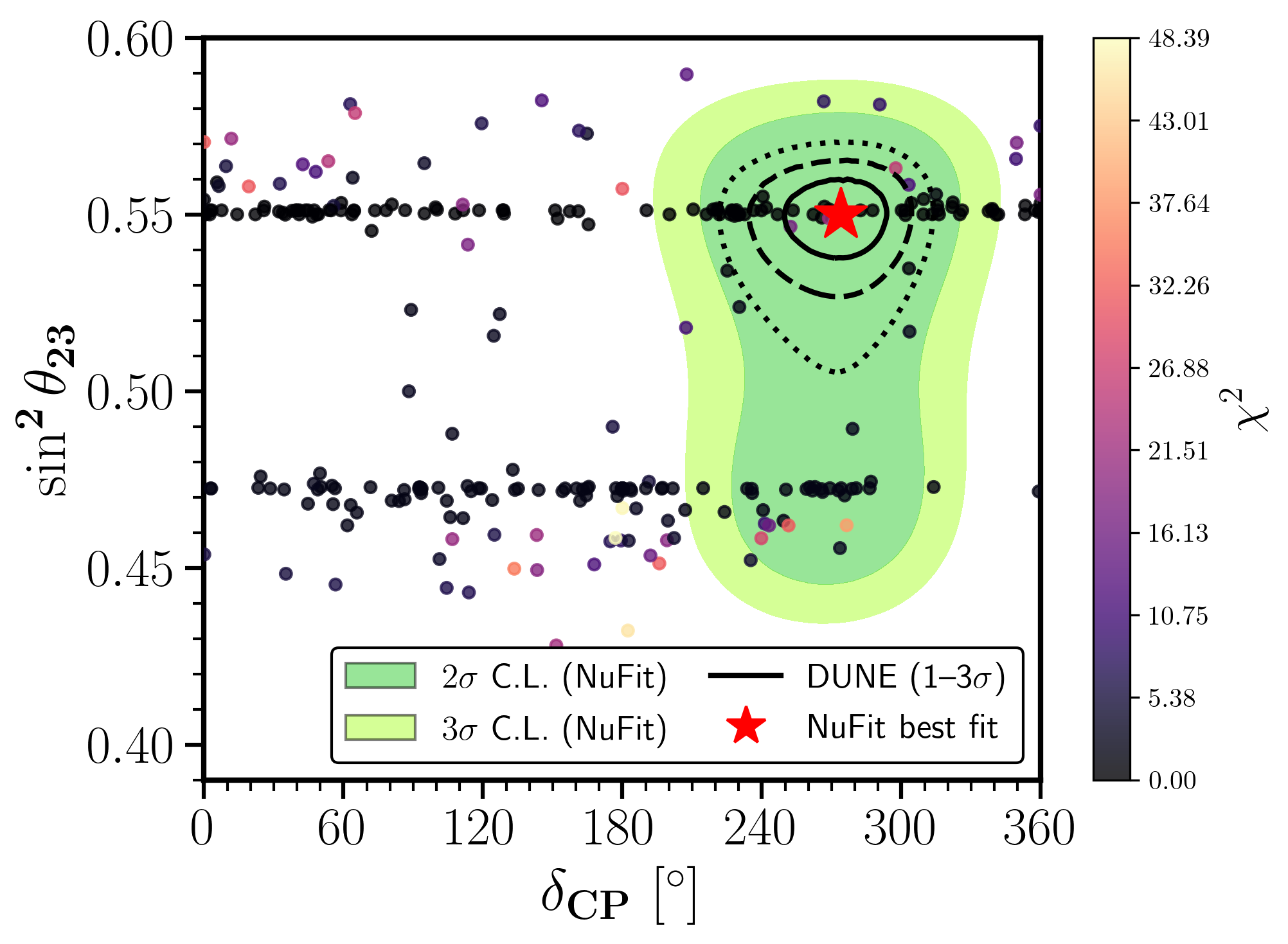}
\caption{    parameter spaces between $\sin^2 \theta_{23}$, $\Delta m^2_{31}/\Delta m^2_{21}$,  and $\delta_{CP}$ by the model representing NO (in left panel) and IO (in right panel). The contour lines denote the allowed regions from the DUNE experiment. The green contours depict the parameter space from NuFIT, while the scattered points represent the region allowed by the model.}
\label{fig:th23-ldm-dcp}
\end{figure}

Varying the value of $\tau$ within the range given in Table~\ref{tab:best-fit-param}, we obtain the allowed parameter regions among $\sin^2 \theta_{13}$, $\sin^2 \theta_{23}$, $\Delta m^2_{21}$, $\Delta m^2_{31}$, and $\delta_{CP}$, as shown in Figs.~\ref{fig:allowed-no-io} and \ref{fig:th23-ldm-dcp}. The left panels correspond to the NO, while the right panels correspond to the IO. The scattered points represent the parameter space allowed by the model, and the color bars indicate the corresponding $\chi^2$ values of the data points. The $2\sigma$ and $3\sigma$ NuFIT regions are indicated by the green shaded bands. The solid and dashed contours depict the allowed parameter space of the DUNE experiment. Here, we compare the parameter space predicted by the model with the parameter-constrained regions expected from DUNE. Our model allows $\delta_{CP}$ within the full range $[0, 360^\circ]$ and $\sin^2 \theta_{23}$ within $[0.44, 0.60]$. The predicted values of $\sin^2 \theta_{13}$, $\Delta m^2_{21}$, and $\Delta m^2_{31}$ lie within the region allowed by NuFIT.

The allowed regions for the DUNE experiment shown in Fig.~\ref{fig:allowed-no-io} are obtained by minimizing over the oscillation parameters $\theta_{23}$, $\Delta m^2_{31}$, and $\delta_{CP}$, except when the corresponding parameter space is explicitly displayed. From Fig.~\ref{fig:allowed-no-io}, it is evident that DUNE will strongly constrain $\theta_{23}$, $\delta_{CP}$, and $\Delta m^2_{31}$, whereas it has limited sensitivity to $\theta_{13}$. The sensitivity to $\theta_{13}$ at DUNE is affected by parameter degeneracies. In contrast, reactor experiments such as Daya Bay, RENO, and Double Chooz provide high-precision measurements of $\theta_{13}$.

The model constrains $\sin^2 \theta_{13}$ to lie in the range $0.022$–$0.024$ for NO and $0.020$–$0.024$ for IO. The parameter space predicted by the model in the $\Delta m^2_{31}$–$\sin^2 \theta_{13}$ plane (upper panel of Fig.~\ref{fig:allowed-no-io}) lies within the $3\sigma$ allowed regions of both the NuFIT data and the DUNE experiment for both NO and IO scenarios. The parameter space in the $\sin^2 \theta_{13}$–$\sin^2 \theta_{23}$ plane exhibits interesting features: for NO, $\sin^2 \theta_{23}$ shows degenerate solutions around $\sim 0.47$ and $\sim 0.56$. In the IO scenario, the allowed region includes additional parameter space around $\sin^2 \theta_{23} \sim 0.46$ and $\sim 0.55$. Furthermore, the model permits the full range of $\delta_{CP}$ values within $[0, 360^\circ]$.

Fig.~\ref{fig:th23-ldm-dcp} illustrates the allowed parameter space in the $\sin^2\theta_{23}$, $\delta_{\rm CP}$, and the mass-squared difference ($\Delta m^2_{31}$ \& $\Delta m^2_{21}$) planes. The NuFIT best-fit values are indicated by asterisks. The left panels correspond to NO, while the right panels represent IO. The upper panels display the allowed regions in the $\Delta m^2_{31}$–$\delta_{\rm CP}$ plane. For both NO and IO, the model accommodates $\delta_{\rm CP}$ across the full interval $[0,360^\circ]$, while $\Delta m^2_{31}$ is restricted to approximately $[2.46 - 2.50]\times10^{-3}\mathrm{~eV}^2$. The model-predicted parameter space lies within the $3\sigma$ region allowed by NuFIT. The projected sensitivity of DUNE significantly constrains the $\Delta m^2_{31}$–$\delta_{\rm CP}$ parameter space, and a subset of the model-preferred points lie within the $3\sigma$ contour of DUNE experiment.

The  panels in the second and third rows show the allowed regions in the $\Delta m^2_{31}$–$\sin^2\theta_{23}$ and $\Delta m^2_{21}$–$\sin^2\theta_{23}$ planes. The preferred ranges obtained from the model are consistent with the global oscillation analysis of NuFIT.   The solar mass squared difference $\Delta m^2_{21}$ is constrained to a narrow range, approximately $[7.4 -7.6 ]\times10^{-5}\mathrm{~eV}^2$. Although, $\sin^2\theta_{23}$ spans the interval $[0.4,0.6]$, the model exhibits a pronounced preference near $\sin^2\theta_{23}\simeq0.47$ for NO and $\sin^2\theta_{23}\simeq0.55$ for IO. Degenerate solutions are observed in both orderings, reflecting the well-known octant degeneracy. The spread of the NuFIT-allowed region corresponds to the coexistence of lower and higher-octant solutions. The DUNE sensitivity projections in the $\Delta m^2_{31}$–$\sin^2\theta_{23}$ plane demonstrate a substantial reduction of the allowed parameter space, indicating its capability to resolve the octant–hierarchy degeneracy. A non-negligible fraction of the model predictions overlaps with the $3\sigma$ DUNE sensitivity region.  Since  DUNE is not sensitive enough to the  solar oscillation parameters, we have not explicitly shown the DUNE sensitivity contours in the $\Delta m^2_{21}$–$\sin^2\theta_{23}$ plane.

The plots in the bottom  panel present the allowed parameter space in the $\delta_{\rm CP}$–$\sin^2\theta_{23}$ plane. The model permits $\delta_{\rm CP}$ throughout $[0,360^\circ]$ and $\sin^2\theta_{23}$ within $[0.4,0.6]$, with a densely populated region  around $\sin^2\theta_{23}\approx0.47$ and $0.55$. In contrast, the DUNE projections indicate a significantly narrower allowed region. Overall, the model predictions are compatible with the current NuFIT constraints and partially fall within the projected sensitivity reach of DUNE.

\subsection{Neutrinoless Double Beta Decay} 
\label{sec:NDBD}
In this section, we have shown the model's capability to predict the effective electron neutrino mass, the effective Majorana neutrino mass, and the sum of neutrino masses. 

\begin{figure}
\centering
\includegraphics[width=.48\textwidth]{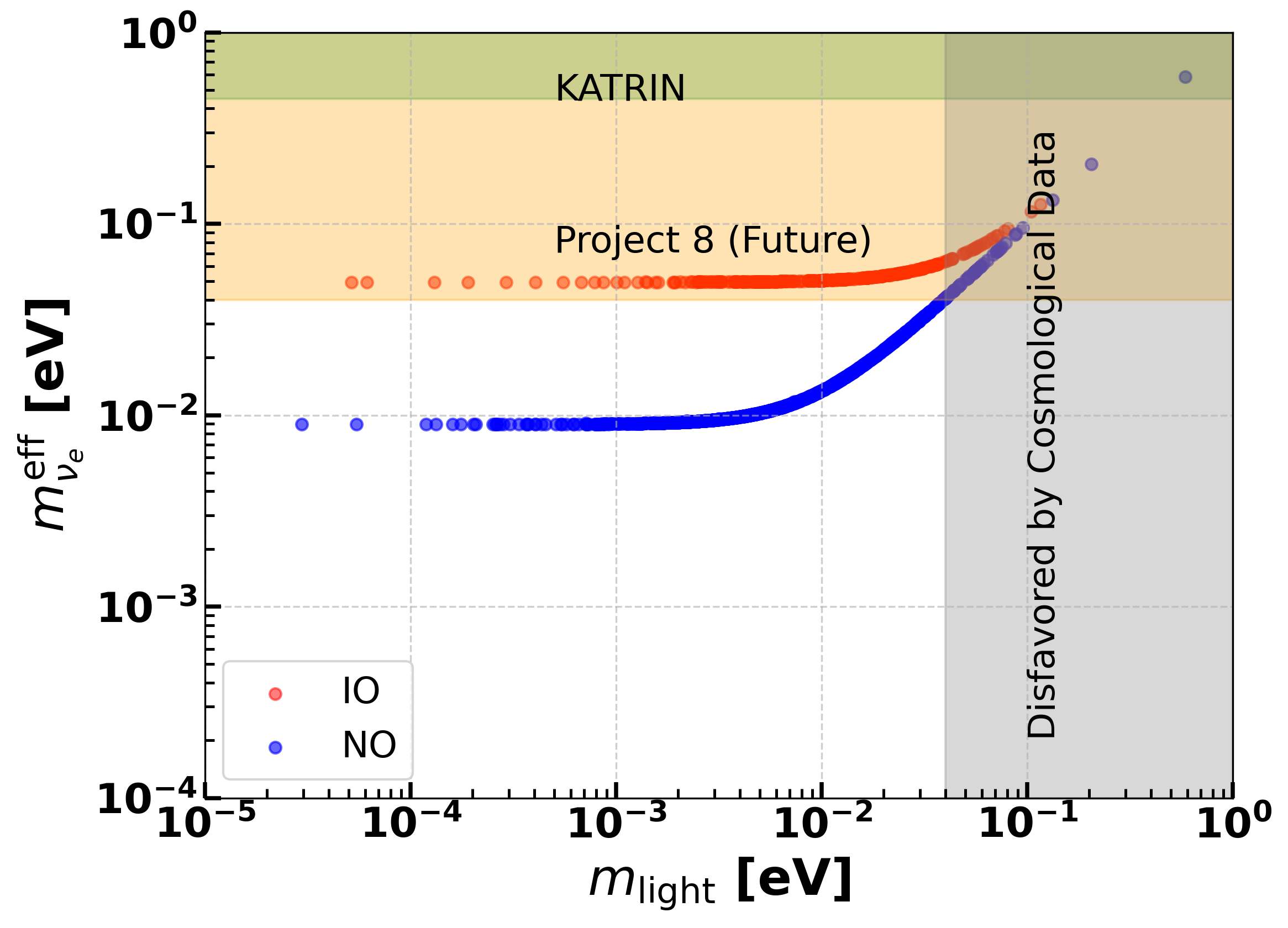}
\includegraphics[width=.48\textwidth]{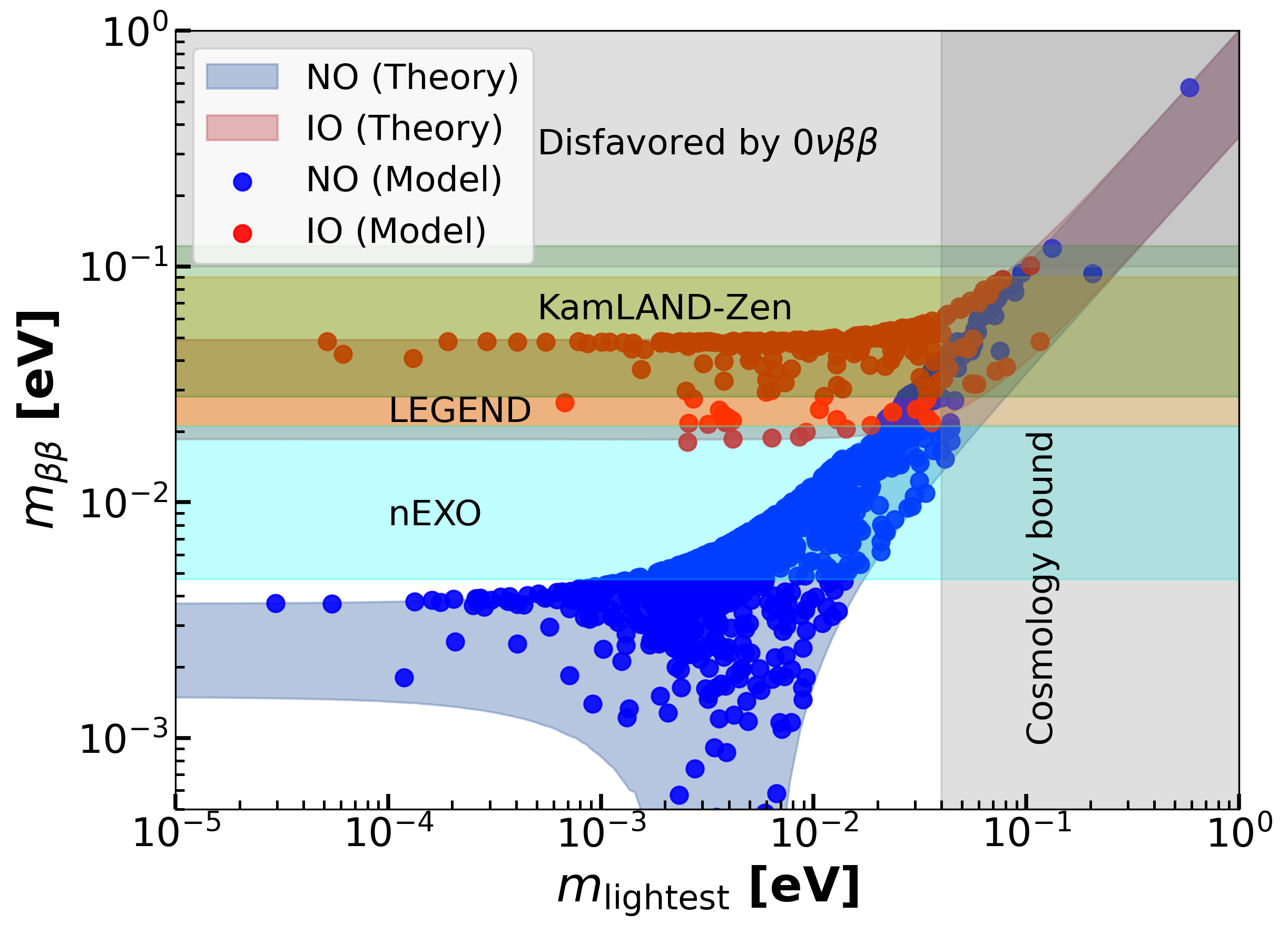}
\caption{Effective Majorana mass for neutrino $m_{\beta\beta}$ (right panel) and $m_\nu^{\rm eff}$ (left panel) as a function of lightest neutrino mass for NO and IO, allowed by the model.}
\label{fig:mbb-mnu-mlight}
\end{figure}

The effective mass of the electron neutrino can be determined from the endpoint energy of the beta decay spectrum and can be expressed as 
\begin{equation}
m^{\rm eff}_{\nu_e} = \sqrt{\sum_i |U_{ei}|^2 m_{i}^2}
\end{equation}
where $U_{ei}$ is the mixing matrix elements and $ m_{i}$ is the mass of $i^{\rm th}$ neutrino mass eigenstate. The left panel of Fig. \ref{fig:mbb-mnu-mlight} shows the effective neutrino mass $m^{\rm eff}_{\nu_e}$ predicted by the model for NO and IO, with red and blue points, respectively. The gray shaded region shows the region disfavored by cosmological data. Experimental data from  KATRIN  put constraints on $m^{\rm eff}_{\nu_e}  < 0.45$ eV at $90\%$ C.L. \cite{KATRIN:2024cdt}. Project 8 can also put very stringent upper bound on $m^{\rm eff}_{\nu_e} < 0.04 \mathrm{~eV}$ \cite{Project8:2022wqh}. In the NO case, the entire model-predicted parameter space for $m^{\rm eff}_{\nu_e} \simeq 0.008\text{--}0.015~\text{eV}$ remains below the expected sensitivity of the Project~8 experiment. For the IO case, the predicted model parameter space $m^{\rm eff}_{\nu_e} \simeq 0.045\text{--}0.060~\text{eV}$ lies above the Project~8 sensitivity limit.

Neutrinoless double beta decay ($0\nu\beta\beta$) may shed light on the Majorana nature of neutrinos. The half-life of $0\nu\beta\beta$ process can be expressed as 
\begin{equation}
T^{-1}_{1/2} \simeq {G_{0\nu}}\left | \frac{{\cal M}_{N}}{m_e} \right |^2  |m_{\beta\beta}|^2 ,
\end{equation}
where $G_{0\nu}$ is phase space factor, $m_e$ is the mass of electron, ${\cal M}_N$ is the nuclear matrix element for the nucleus, and $|m_{\beta\beta}|$ is the absolute effective Majorana mass parameter. The effective Majorana mass parameter can be characterized in terms of the PMNS mixing matrix elements as 
\begin{equation}
|m_{\beta\beta}| = |m_1 |U_{e1}|^2 + m_2 |U_{e2}|^2 e^{i\alpha_{21}}+ m_3 |U_{e3}|^2 e^{i(\alpha_{31}-2\delta_{CP})}  |,
\end{equation}
where the $m_i$'s are the neutrino masses and $U_{ei}$'s are the mixing matrix elements associated with electron with ($i=1,2,3$), which depend upon the mixing angles. Phases $\alpha_{21}$ and $\alpha_{31}$ are the Majorana phases and $\delta_{CP}$ is the Dirac CP phase. Many dedicated experiments are looking for neutrinoless double beta signals; for details, please refer to \cite{Giuliani:2019uno}. The sensitivity limits  on $\left | m_{\beta \beta}\right |$ by the current experiments  such as GERDA  is $(102-213)$ meV \cite{Agostini:2019hzm} and CUORE is $(90-420)$ meV  \cite{Alduino:2017ehq}. The recent result of KamLAND-Zen experiment constrains  $m_{\beta \beta}<(28-122)$ meV \cite{KamLAND-Zen:2024eml}. 
The future generation experiments, like LEGEND-1000 can probe $9-21$ meV \cite{LEGEND:2021bnm} and nEXO can reach $4.7-20.3$ meV in the future \cite{nEXO:2021ujk}.  In Fig.~\ref{fig:mbb-mnu-mlight} (right panel), we have shown the $m_{\beta\beta}$ as a function of lightest neutrino mass. We present the predictions of our model as data points in blue and red for the NO and IO cases, respectively. The light-blue and light-red shaded regions represent the theoretical allowed NO and IO parameter space, respectively. The horizontal gray shaded band corresponds to the region disfavored by $0\nu\beta\beta$ searches, while the vertical shaded band indicates the region excluded by cosmological constraints.

For both the mass ordering cases, the majority of the parameter space predicted by the model falls within the corresponding theoretical region. 
For NO, the predicted values of $m_{\beta\beta} \simeq 0.0008\text{--}0.04~\text{eV}$  lie below the present experimental bounds from LEGEND and KamLAND-Zen, although a part of the parameter space remains within the projected sensitivity of the nEXO experiment. In contrast, for IO, the predicted values of  $m_{\beta\beta} \simeq 0.017\text{--}0.061~\text{eV}$ fall within the future discovery reach of nEXO and LEGEND, while still remaining below the current observational sensitivity of KamLAND-Zen experiment.

\begin{figure}[htpb]
\centering
\includegraphics[width=.48\textwidth]{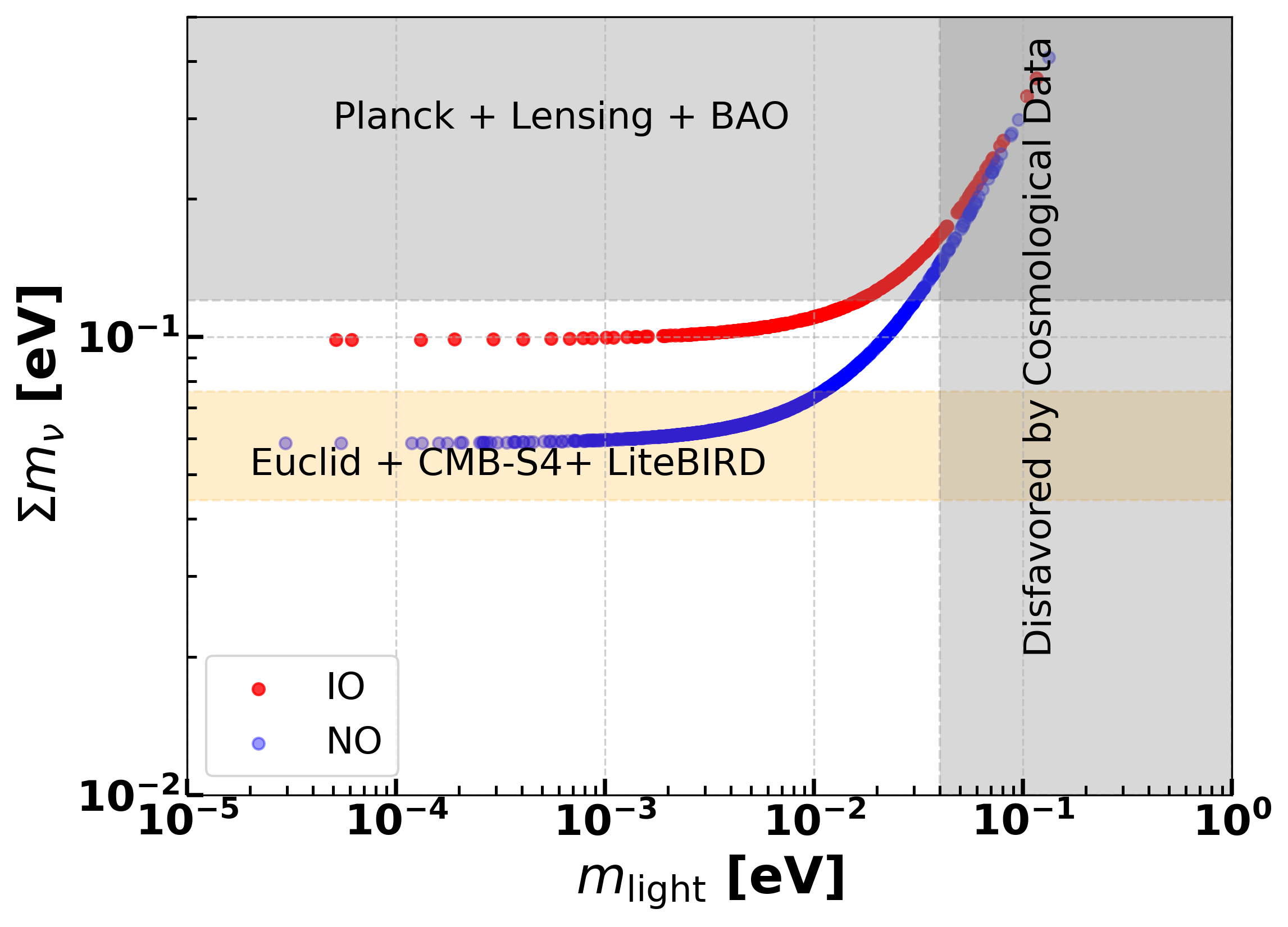}
\caption{Sum of neutrino masses $\sum m_\nu$ as a function of lightest neutrino mass for NO and IO, allowed by the model.}
\label{fig:sum-mnu-mlight}
\end{figure}

Figure~\ref{fig:sum-mnu-mlight} shows the sum of neutrino masses as a function of the lightest neutrino mass. The Planck measurement of the cosmic microwave background (CMB) for cosmological parameters provides a stringent constraint on the sum of neutrino masses, $\sum m_\nu < 0.12$ eV \cite{Planck:2018vyg}. A recent result from the DESI Collaboration reports that the combined analysis of DESI BAO and CMB data sets a tighter upper bound, $\sum m_\nu < 0.072$ eV \cite{DESI:2024mwx}. In the future, the joint analysis of Euclid, CMB$-$S4, and LiteBIRD data is expected to measure $\sum m_\nu$ in the range ($0.044 -0.076$) eV \cite{Euclid:2024imf}. The model prediction for $\sum m_\nu$ for the NO scenario lies below the upper limit of Planck + Lensing + BAO data. Most of the predicted data in IO case is above the upper limit of Planck + Lensing + BAO data. However, the prediction in the case of IO lie in $\sum m_\nu \simeq 0.10\text{--}0.16~\text{eV}$,  exceeds the futuristic upper limit of   Euclid, CMB$-$S4, and LiteBIRD experiments. The range of prediction for $\sum m_\nu \simeq 0.058\text{--}0.13~\text{eV}$ in NO is within the expected sensitivity range of Euclid+ CMB$-$S4, + LiteBIRD limit.
\section{Summary and Conclusions}
\label{sec: summary}
In this work, we have presented a predictive realization of the linear seesaw mechanism within a non-holomorphic modular $A_4$ flavor symmetry framework using polyharmonic Maaß modular forms in a non-supersymmetric setup. The model introduces six gauge-singlet fermions $(N_{Ri}, S_{Li})$ together with a single flavon field, thereby significantly reducing scalar-sector arbitrariness compared to conventional $A_4$ constructions and earlier holomorphic modular linear seesaw models. Once the modulus $\tau$ acquires a vacuum expectation value in the fundamental domain, the entire flavor structure is dynamically fixed. Under the hierarchy $M_{LS} \ll M_D \ll M_{RS}$ with $M_N \ll M_{RS}$, the effective light neutrino mass matrix naturally assumes the linear seesaw structure with dominant scaling $m_\nu \sim M_D M_{LS}/M_{RS}$. Our numerical scan reveals viable solutions for $\text{Re}(\tau) \in [-0.5,\,0.5]$ and $\text{Im}(\tau) \in [0.86,\,5]$ in the NO case, while the IO solutions are more restricted, clustering around $\text{Re}(\tau)\approx 0$ and $\text{Im}(\tau)$ spans in a wider parameter space. For NO, the model predicts $\sin^2\theta_{12} \simeq 0.27\text{--}0.35$, $\sin^2\theta_{13} \simeq 0.020\text{--}0.024$, and $\sin^2\theta_{23} \simeq 0.46\text{--}0.48$ or $0.53\text{--}0.56$, clearly exhibiting an octant degeneracy. The atmospheric mass squared differences lie in the range of $|\Delta m_{31}^2| \simeq (2.45\text{--}2.5)\times 10^{-3}\,\text{eV}^2$. In the IO scenario, the allowed region is comparatively compressed around $\sin^2\theta_{23} \simeq 0.47\text{ and }0.55$, and $|\Delta m_{31}^2| \simeq 2.46\times 10^{-3}\,\text{eV}^2$, while the Dirac CP phase $\delta_{\rm CP}$ spans nearly the full physical range $[0,2\pi]$. The solar mass squared difference $\Delta m^2_{21}$ is constrained to a narrow range, approximately $[7.4 -7.6 ]\times10^{-5}\mathrm{~eV}^2$. In the absolute neutrino mass sector, we obtain $\sum m_\nu \simeq 0.058\text{--}0.13~\text{eV}$ for NO and $\sum m_\nu \simeq 0.10\text{--}0.16~\text{eV}$ for IO, placing much of the IO parameter space close to present cosmological bounds. The effective electron neutrino mass is predicted to be $m^{\rm eff}_{\nu_e} \simeq 0.008\text{--}0.015~\text{eV}$ in NO and $m^{\rm eff}_{\nu_e} \simeq 0.045\text{--}0.060~\text{eV}$ in IO. The effective Majorana mass relevant for neutrinoless double beta decay lies in the ranges $m_{\beta\beta} \simeq 0.0008\text{--}0.04~\text{eV}$ for NO and $m_{\beta\beta} \simeq 0.017\text{--}0.061~\text{eV}$ for IO, with the IO region significantly overlapping the projected sensitivities of next-generation experiments.
Furthermore, our GLoBES-based simulation of DUNE indicates that future precision measurements will substantially reduce the allowed parameter space in the $(\theta_{23},\delta_{\rm CP})$ and $(\Delta m_{31}^2,\delta_{\rm CP})$ planes, potentially resolving the octant degeneracy predicted by the model. Overall, the non-holomorphic modular $A_4$ linear seesaw framework provides a theoretically economical and symmetry-driven description of neutrino mass generation, yielding sharply defined numerical predictions and distinctive parameter correlations that can be tested in upcoming oscillation, cosmological, and neutrinoless double beta decay experiments.

\appendix
\section{Modular Yukawa couplings}
\label{app:A}
The functions $Y^{(0)}_{3,i}$ and $Y^{(-2)}_{3,i}$ $(i=1,2,3)$ are modular Yukawa couplings transforming as an $A_4$ triplet with modular weights $0$ and $-2$, respectively, expressed in terms of the modulus $\tau = x + i y$ and $q = e^{2\pi i \tau}$. Their $q$-expansions are given as follows:
\begin{align}
Y^{(0)}_{3,1} &= y
- \frac{3 e^{-4\pi y}}{\pi q}
- \frac{9 e^{-8\pi y}}{2\pi q^{2}}
- \frac{e^{-12\pi y}}{\pi q^{3}}
- \frac{21 e^{-16\pi y}}{4\pi q^{4}}
- \frac{18 e^{-20\pi y}}{5\pi q^{5}}
- \frac{3 e^{-24\pi y}}{2\pi q^{6}}
+ \cdots \nonumber\\
&\quad
- \frac{9 \log 3}{4\pi}
- \frac{3q}{2\pi}
- \frac{9q^{2}}{2\pi}
- \frac{q^{3}}{\pi}
- \frac{21q^{4}}{4\pi}
- \frac{18q^{5}}{5\pi}
- \frac{3q^{6}}{2\pi}
+ \cdots ,
\label{A-1}
\end{align}
\begin{align}
Y^{(0)}_{3,2} &= 
\frac{27 q^{1/3} e^{\pi y/3}}{\pi}
\left(
\frac{e^{-3\pi y}}{4q}
+ \frac{e^{-7\pi y}}{5q^{2}}
+ \frac{5e^{-11\pi y}}{16q^{3}}
+ \frac{2e^{-15\pi y}}{11q^{4}}
+ \frac{2e^{-19\pi y}}{7q^{5}}
+ \frac{4e^{-23\pi y}}{17q^{6}}
+ \cdots
\right)
\nonumber\\
&\quad
+ \frac{9 q^{1/3}}{2\pi}
\left(
1 + \frac{7q}{4}
+ \frac{8q^{2}}{7}
+ \frac{9q^{3}}{5}
+ \frac{14q^{4}}{13}
+ \frac{31q^{5}}{16}
+ \frac{20q^{6}}{19}
+ \cdots
\right),
\label{A-2}
\end{align}
\begin{align}
Y^{(0)}_{3,3} &= 
\frac{9 q^{2/3} e^{2\pi y/3}}{2\pi}
\left(
\frac{e^{-2\pi y}}{q}
+ \frac{7e^{-6\pi y}}{4q^{2}}
+ \frac{8e^{-10\pi y}}{7q^{3}}
+ \frac{9e^{-14\pi y}}{5q^{4}}
+ \frac{14e^{-18\pi y}}{13q^{5}}
+ \frac{31e^{-22\pi y}}{16q^{6}}
+ \cdots
\right)
\nonumber\\
&\quad
+ \frac{27 q^{2/3}}{\pi}
\left(
\frac{1}{4}
+ \frac{q}{5}
+ \frac{5q^{2}}{16}
+ \frac{2q^{3}}{11}
+ \frac{2q^{4}}{7}
+ \frac{9q^{5}}{17}
+ \frac{21q^{6}}{20}
+ \cdots
\right).
\label{A-3}
\end{align}

\begin{align}
Y^{(-2)}_{3,1}(\tau)
=&\;
\frac{y^3}{3}
+\frac{21\,\Gamma(3,4\pi y)}{16\pi^3\,q}
+\frac{189\,\Gamma(3,8\pi y)}{128\pi^3\,q^2}
+\frac{169\,\Gamma(3,12\pi y)}{144\pi^3\,q^3}
+\frac{1533\,\Gamma(3,16\pi y)}{1024\pi^3\,q^4}
+\cdots
\nonumber\\[2mm]
&\;
+\frac{\pi\,\zeta(3)}{40\,\zeta(4)}
+\frac{21\,q}{8\pi^3}
+\frac{189\,q^2}{64\pi^3}
+\frac{169\,q^3}{72\pi^3}
+\frac{1533\,q^4}{512\pi^3}
+\frac{1323\,q^5}{500\pi^3}
+\frac{169\,q^6}{64\pi^3}
+\cdots ,
\label{A-4}
\end{align}
\begin{align}
Y^{(-2)}_{3,2}(\tau)
=&\;
-\frac{729\,q^{1/3}}{16\pi^3}
\left(
\frac{\Gamma(3,8\pi y/3)}{16\,q}
+\frac{7\,\Gamma(3,20\pi y/3)}{125\,q^2}
+\frac{65\,\Gamma(3,32\pi y/3)}{1024\,q^3}
+\frac{74\,\Gamma(3,44\pi y/3)}{1331\,q^4}
+\cdots
\right)
\nonumber\\[2mm]
&\;
-\frac{81\,q^{1/3}}{16\pi^3}
\left(
1
+\frac{73\,q}{64}
+\frac{344\,q^2}{343}
+\frac{567\,q^3}{500}
+\frac{20198\,q^4}{2197}
+\frac{4681\,q^5}{4096}
+\cdots
\right) ,
\label{A-5}
\end{align}
\begin{align}
Y^{(-2)}_{3,3}(\tau)
=&\;
-\frac{81\,q^{2/3}}{32\pi^3}
\left(
\frac{\Gamma(3,4\pi y/3)}{q}
+\frac{73\,\Gamma(3,16\pi y/3)}{64\,q^2}
+\frac{344\,\Gamma(3,28\pi y/3)}{343\,q^3}
+\frac{567\,\Gamma(3,40\pi y/3)}{500\,q^4}
+\cdots
\right)
\nonumber\\[2mm]
&\;
-\frac{729\,q^{2/3}}{8\pi^3}
\left(
\frac{1}{16}
+\frac{7\,q}{125}
+\frac{65\,q^2}{1024}
+\frac{74\,q^3}{1331}
+\cdots
\right) .
\label{A-6}
\end{align}

\section{Block Diagonalization}
\label{appex-B}
For a block matrix $M$  
\begin{equation}
    M = \left( \begin{array}{cc}
        A & B \\
         B^T & D
    \end{array}
    \right),
\end{equation}
where the $A$, $B$, and $D$ are $a\times a$, $a\times b$, and $b\times b$  matrices. The dimension of $M$ is $(a+b)\times (a+b)$. Schur complement \cite{Gallier2010Schur} of the block $D$ when $D$ is invertible is  
\begin{equation}
    S_D = A - B D^{-1} B^T.
    \label{schur}
\end{equation}

In this study, the mass matrix is in the basis of $(\nu_l, N_R, S_L^c)$ basis can be written as 
\begin{equation}
    M = \left(\begin{array}{ccc}
        0 & M_D & M_{LS}  \\
         M_D ^T& M_N & M_{RS} \\
         M_{LS}^T & M_{RS}^T & 0
    \end{array}\right).
\end{equation}
The mass matrix $M$ can be block diagonalized through Schur complement using realizing 
\begin{eqnarray}\nonumber
    A = 0,\\ \nonumber
    B = \left( M_D~~ M_{LS}\right) \\
    D = \left( \begin{array}{cc}
        M_N & M_{RS} \\
        M_{RS}^T & 0
    \end{array}
    \right),
\end{eqnarray}
and the Schur complement as 
\begin{eqnarray}
        S_D & = & - \left( \begin{array}{cc}
        M_D & M_{LS} 
    \end{array}
    \right) \left( \begin{array}{cc}
        0 & ({M_{RS}}^T)^{-1} \\
        M_{RS}^{-1} & - M_{RS}^{-1} M_N ({M_{RS}}^T)^{-1}
    \end{array}
    \right) \left( \begin{array}{c}
        M_D^T \\ M_{LS}^T 
    \end{array}
    \right) \\
    m_\nu   & =& M_D (M_{LS}M_{RS}^{-1})^T + (M_{LS}M_{RS}^{-1}) M_D^T - M_{LS}M_{RS}^{-1}M_N ({M_{RS}^T})^{-1}M_{LS}^T 
\end{eqnarray}
\acknowledgments

RM (Rudra Majhi) would like to acknowledge Odisha State Higher Education Council, Govt. of Odisha for the support under Mukhyamantri Research and Innovation (MRIP)-2024 (24EM/PH/102).  
\bibliographystyle{JHEP}
\bibliography{biblio}

@article{DUNE:2020jqi,
    author = "Abi, B. and others",
    collaboration = "DUNE",
    title = "{Long-baseline neutrino oscillation physics potential of the DUNE experiment}",
    eprint = "2006.16043",
    archivePrefix = "arXiv",
    primaryClass = "hep-ex",
    reportNumber = "FERMILAB-PUB-20-251-E-LBNF-ND-PIP2-SCD, PUB-20-251-E-LBNF-ND-PIP2-SCD",
    doi = "10.1140/epjc/s10052-020-08456-z",
    journal = "Eur. Phys. J. C",
    volume = "80",
    number = "10",
    pages = "978",
    year = "2020"
}

@article{deSalas:2020pgw,
    author = "de Salas, P. F. and Forero, D. V. and Gariazzo, S. and Mart{\'\i}nez-Mirav{\'e}, P. and Mena, O. and Ternes, C. A. and T{\'o}rtola, M. and Valle, J. W. F.",
    title = "{2020 global reassessment of the neutrino oscillation picture}",
    eprint = "2006.11237",
    archivePrefix = "arXiv",
    primaryClass = "hep-ph",
    doi = "10.1007/JHEP02(2021)071",
    journal = "JHEP",
    volume = "02",
    pages = "071",
    year = "2021"
}

@article{Capozzi:2021fjo,
    author = "Capozzi, Francesco and Di Valentino, Eleonora and Lisi, Eligio and Marrone, Antonio and Melchiorri, Alessandro and Palazzo, Antonio",
    title = "{Unfinished fabric of the three neutrino paradigm}",
    eprint = "2107.00532",
    archivePrefix = "arXiv",
    primaryClass = "hep-ph",
    doi = "10.1103/PhysRevD.104.083031",
    journal = "Phys. Rev. D",
    volume = "104",
    number = "8",
    pages = "083031",
    year = "2021"
}

@article{KATRIN:2024cdt,
    author = "Aker, Max and others",
    collaboration = "KATRIN",
    title = "{Direct neutrino-mass measurement based on 259 days of KATRIN data}",
    eprint = "2406.13516",
    archivePrefix = "arXiv",
    primaryClass = "nucl-ex",
    doi = "10.1126/science.adq9592",
    journal = "Science",
    volume = "388",
    number = "6743",
    pages = "adq9592",
    year = "2025"
}

@article{Kashav:2022kpk,
    author = "Kashav, Monal and Verma, Surender",
    title = "{On minimal realization of topological Lorentz structures with one-loop seesaw extensions in A$_{4}$ modular symmetry}",
    eprint = "2205.06545",
    archivePrefix = "arXiv",
    primaryClass = "hep-ph",
    doi = "10.1088/1475-7516/2023/03/010",
    journal = "JCAP",
    volume = "03",
    pages = "010",
    year = "2023"
}

@article{Kumar:2023moh,
    author = "Kumar, Ranjeet and Mishra, Priya and Behera, Mitesh Kumar and Mohanta, Rukmani and Srivastava, Rahul",
    title = "{Predictions from scoto-seesaw with A4 modular symmetry}",
    eprint = "2310.02363",
    archivePrefix = "arXiv",
    primaryClass = "hep-ph",
    doi = "10.1016/j.physletb.2024.138635",
    journal = "Phys. Lett. B",
    volume = "853",
    pages = "138635",
    year = "2024"
}

@article{Behera:2020lpd,
    author = "Behera, Mitesh Kumar and Singirala, Shivaramakrishna and Mishra, Subhasmita and Mohanta, Rukmani",
    title = "{A modular A $_{4}$ symmetric scotogenic model for neutrino mass and dark matter}",
    eprint = "2009.01806",
    archivePrefix = "arXiv",
    primaryClass = "hep-ph",
    doi = "10.1088/1361-6471/ac3cc2",
    journal = "J. Phys. G",
    volume = "49",
    number = "3",
    pages = "035002",
    year = "2022"
}

@article{Behera:2020sfe,
    author = "Behera, Mitesh Kumar and Mishra, Subhasmita and Singirala, Shivaramakrishna and Mohanta, Rukmani",
    title = "{Implications of A4 modular symmetry on neutrino mass, mixing and leptogenesis with linear seesaw}",
    eprint = "2007.00545",
    archivePrefix = "arXiv",
    primaryClass = "hep-ph",
    doi = "10.1016/j.dark.2022.101027",
    journal = "Phys. Dark Univ.",
    volume = "36",
    pages = "101027",
    year = "2022"
}

@article{Nomura:2023kwz,
    author = "Nomura, Takaaki and Okada, Hiroshi",
    title = "{Quark and lepton model with flavor specific dark matter and muon g{\ensuremath{-}}2 in modular A4 and hidden U(1) symmetries}",
    eprint = "2304.13361",
    archivePrefix = "arXiv",
    primaryClass = "hep-ph",
    doi = "10.1016/j.dark.2025.101986",
    journal = "Phys. Dark Univ.",
    volume = "49",
    pages = "101986",
    year = "2025"
}

@article{Kim:2023jto,
    author = "Kim, Jongkuk and Okada, Hiroshi",
    title = "{Fermi-LAT GeV excess and muon $g-2$ in a modular $A_4$ symmetry}",
    eprint = "2302.09747",
    archivePrefix = "arXiv",
    primaryClass = "hep-ph",
    month = "2",
    year = "2023"
}

@article{Devi:2023vpe,
    author = "Devi, Maibam Ricky",
    title = "{Retrieving texture zeros in 3+1 active-sterile neutrino framework under the action of $A_4$ modular-invariants}",
    eprint = "2303.04900",
    archivePrefix = "arXiv",
    primaryClass = "hep-ph",
    month = "3",
    year = "2023"
}

@article{Dasgupta:2021ggp,
    author = "Dasgupta, Arnab and Nomura, Takaaki and Okada, Hiroshi and Popov, Oleg and Tanimoto, Morimitsu",
    title = "{Dirac Radiative Neutrino Mass with Modular Symmetry and Leptogenesis}",
    eprint = "2111.06898",
    archivePrefix = "arXiv",
    primaryClass = "hep-ph",
    reportNumber = "APCTP Pre2021 - 029, CTP-SCU/2021033",
    month = "11",
    year = "2021"
}

@article{Nomura:2019lnr,
    author = "Nomura, Takaaki and Okada, Hiroshi and Popov, Oleg",
    title = "{A modular $A_4$ symmetric scotogenic model}",
    eprint = "1908.07457",
    archivePrefix = "arXiv",
    primaryClass = "hep-ph",
    reportNumber = "KIAS-P19048, APCTP Pre2019 - 022",
    doi = "10.1016/j.physletb.2020.135294",
    journal = "Phys. Lett. B",
    volume = "803",
    pages = "135294",
    year = "2020"
}

@article{Mishra:2022egy,
    author = "Mishra, Priya and Behera, Mitesh Kumar and Panda, Papia and Mohanta, Rukmani",
    title = "{Type III seesaw under $A_4$ modular symmetry with leptogenesis}",
    eprint = "2204.08338",
    archivePrefix = "arXiv",
    primaryClass = "hep-ph",
    doi = "10.1140/epjc/s10052-022-11074-6",
    journal = "Eur. Phys. J. C",
    volume = "82",
    number = "12",
    pages = "1115",
    year = "2022"
}

@article{CentellesChulia:2023osj,
    author = "Centelles Chuli{\'a}, Salvador and Kumar, Ranjeet and Popov, Oleg and Srivastava, Rahul",
    title = "{Neutrino mass sum rules from modular A4 symmetry}",
    eprint = "2308.08981",
    archivePrefix = "arXiv",
    primaryClass = "hep-ph",
    doi = "10.1103/PhysRevD.109.035016",
    journal = "Phys. Rev. D",
    volume = "109",
    number = "3",
    pages = "035016",
    year = "2024"
}

@article{Gogoi:2023jzl,
    author = "Gogoi, Jotin and Sarma, Lavina and Das, Mrinal Kumar",
    title = "{Leptogenesis and dark matter in minimal inverse seesaw using $A_4$ modular symmetry}",
    eprint = "2311.09883",
    archivePrefix = "arXiv",
    primaryClass = "hep-ph",
    doi = "10.1140/epjc/s10052-024-13029-5",
    journal = "Eur. Phys. J. C",
    volume = "84",
    number = "7",
    pages = "689",
    year = "2024"
}

@article{Pathak:2024sei,
    author = "Pathak, Gourab and Das, Pritam and Das, Mrinal Kumar",
    title = "{Neutrino mass genesis in scoto-inverse seesaw with modular $A_4$}",
    eprint = "2411.13895",
    archivePrefix = "arXiv",
    primaryClass = "hep-ph",
    doi = "10.1140/epjc/s10052-025-14263-1",
    journal = "Eur. Phys. J. C",
    volume = "85",
    number = "5",
    pages = "569",
    year = "2025"
}

@article{Planck:2018vyg,
    author = "Aghanim, N. and others",
    collaboration = "Planck",
    title = "{Planck 2018 results. VI. Cosmological parameters}",
    eprint = "1807.06209",
    archivePrefix = "arXiv",
    primaryClass = "astro-ph.CO",
    doi = "10.1051/0004-6361/201833910",
    journal = "Astron. Astrophys.",
    volume = "641",
    pages = "A6",
    year = "2020",
    note = "[Erratum: Astron.Astrophys. 652, C4 (2021)]"
}

@article{Esteban:2024eli,
    author = "Esteban, Ivan and Gonzalez-Garcia, M. C. and Maltoni, Michele and Martinez-Soler, Ivan and Pinheiro, Jo{\~a}o Paulo and Schwetz, Thomas",
    title = "{NuFit-6.0: updated global analysis of three-flavor neutrino oscillations}",
    eprint = "2410.05380",
    archivePrefix = "arXiv",
    primaryClass = "hep-ph",
    reportNumber = "IFT-UAM/CSIC-24-140, YITP-SB-2024-24, IPPP/24/64, IPPP/24/64, IFT-UAM/CSIC-24-140, YITP-SB-2024-24",
    doi = "10.1007/JHEP12(2024)216",
    journal = "JHEP",
    volume = "12",
    pages = "216",
    year = "2024"
}

@article{Huber:2007ji,
    author = "Huber, Patrick and Kopp, Joachim and Lindner, Manfred and Rolinec, Mark and Winter, Walter",
    title = "{New features in the simulation of neutrino oscillation experiments with GLoBES 3.0: General Long Baseline Experiment Simulator}",
    eprint = "hep-ph/0701187",
    archivePrefix = "arXiv",
    reportNumber = "TUM-HEP-656-07",
    doi = "10.1016/j.cpc.2007.05.004",
    journal = "Comput. Phys. Commun.",
    volume = "177",
    pages = "432--438",
    year = "2007"
}

@article{Huber:2004ka,
    author = "Huber, Patrick and Lindner, M. and Winter, W.",
    title = "{Simulation of long-baseline neutrino oscillation experiments with GLoBES (General Long Baseline Experiment Simulator)}",
    eprint = "hep-ph/0407333",
    archivePrefix = "arXiv",
    reportNumber = "TUM-HEP-553-04",
    doi = "10.1016/j.cpc.2005.01.003",
    journal = "Comput. Phys. Commun.",
    volume = "167",
    pages = "195",
    year = "2005"
}

@article{Huber:2002mx,
    author = "Huber, Patrick and Lindner, Manfred and Winter, Walter",
    title = "{Superbeams versus neutrino factories}",
    eprint = "hep-ph/0204352",
    archivePrefix = "arXiv",
    reportNumber = "TUM-HEP-462-02, MPI-PHT-02-15",
    doi = "10.1016/S0550-3213(02)00825-8",
    journal = "Nucl. Phys. B",
    volume = "645",
    pages = "3--48",
    year = "2002"
}

@article{Zhang:2025dsa,
    author = "Zhang, Xianshuo and Reyimuaji, Yakefu",
    title = "{Inverse seesaw model in nonholomorphic modular A4 flavor symmetry}",
    eprint = "2507.06945",
    archivePrefix = "arXiv",
    primaryClass = "hep-ph",
    doi = "10.1103/17p3-bw5r",
    journal = "Phys. Rev. D",
    volume = "112",
    number = "7",
    pages = "075050",
    year = "2025"
}

@article{Liu:2020msy,
    author = "Liu, Xiang-Gan and Yao, Chang-Yuan and Qu, Bu-Yao and Ding, Gui-Jun",
    title = "{Half-integral weight modular forms and application to neutrino mass models}",
    eprint = "2007.13706",
    archivePrefix = "arXiv",
    primaryClass = "hep-ph",
    reportNumber = "USTC-ICTS/PCFT-20-22",
    doi = "10.1103/PhysRevD.102.115035",
    journal = "Phys. Rev. D",
    volume = "102",
    number = "11",
    pages = "115035",
    year = "2020"
}

@article{Fogli:2002pt,
    author = "Fogli, G. L. and Lisi, E. and Marrone, A. and Montanino, D. and Palazzo, A.",
    title = "{Getting the most from the statistical analysis of solar neutrino oscillations}",
    eprint = "hep-ph/0206162",
    archivePrefix = "arXiv",
    doi = "10.1103/PhysRevD.66.053010",
    journal = "Phys. Rev. D",
    volume = "66",
    pages = "053010",
    year = "2002"
}

@article{Pakvasa:1977in,
    author = "Pakvasa, Sandip and Sugawara, Hirotaka",
    title = "{Discrete Symmetry and Cabibbo Angle}",
    reportNumber = "UH-511-240-77",
    doi = "10.1016/0370-2693(78)90172-7",
    journal = "Phys. Lett. B",
    volume = "73",
    pages = "61--64",
    year = "1978"
}

@article{Feruglio:2017ieh,
    author = "Feruglio, Ferruccio",
    editor = "Fusco Femiano, R. and Mannocchi, G. and Morselli, A. and Trinchero, G.",
    title = "{Neutrino masses and mixing angles: A tribute to Guido Altarelli}",
    journal = "Frascati Phys. Ser.",
    volume = "64",
    pages = "174--182",
    year = "2017"
}

@article{Tanabashi:2018oca,
      author         = "Tanabashi, M. and others",
      title          = "{Review of Particle Physics}",
      collaboration  = "Particle Data Group",
      journal        = "Phys. Rev.",
      volume         = "D98",
      year           = "2018",
      number         = "3",
      pages          = "030001",
      doi            = "10.1103/PhysRevD.98.030001",
      SLACcitation   = "%%CITATION = PHRVA,D98,030001;%%"
}

@article{Weinberg:1980bf,
      author         = "Weinberg, Steven",
      title          = "{Varieties of Baryon and Lepton Nonconservation}",
      journal        = "Phys. Rev.",
      volume         = "D22",
      year           = "1980",
      pages          = "1694",
      doi            = "10.1103/PhysRevD.22.1694",
      reportNumber   = "HUTP-80/A023",
      SLACcitation   = "%%CITATION = PHRVA,D22,1694;%%"
}

@article{Weinberg:1979sa,
      author         = "Weinberg, Steven",
      title          = "{Baryon and Lepton Nonconserving Processes}",
      journal        = "Phys. Rev. Lett.",
      volume         = "43",
      year           = "1979",
      pages          = "1566-1570",
      doi            = "10.1103/PhysRevLett.43.1566",
      reportNumber   = "HUTP-79-A050",
      SLACcitation   = "%%CITATION = PRLTA,43,1566;%%"
}

@article{Wilczek:1979hc,
      author         = "Wilczek, Frank and Zee, A.",
      title          = "{Operator Analysis of Nucleon Decay}",
      journal        = "Phys. Rev. Lett.",
      volume         = "43",
      year           = "1979",
      pages          = "1571-1573",
      doi            = "10.1103/PhysRevLett.43.1571",
      reportNumber   = "Print-79-0709 (PRINCETON)",
      SLACcitation   = "%%CITATION = PRLTA,43,1571;%%"
}

@article{Minkowski:1977sc,
      author         = "Minkowski, Peter",
      title          = "{$\mu \to e\gamma$ at a Rate of One Out of $10^{9}$ Muon
                        Decays?}",
      journal        = "Phys. Lett.",
      volume         = "67B",
      year           = "1977",
      pages          = "421-428",
      doi            = "10.1016/0370-2693(77)90435-X",
      reportNumber   = "Print-77-0182 (BERN)",
      SLACcitation   = "%%CITATION = PHLTA,67B,421;%%"
}

@article{Mohapatra:1979ia,
      author         = "Mohapatra, Rabindra N. and Senjanovic, Goran",
      title          = "{Neutrino Mass and Spontaneous Parity Nonconservation}",
      journal        = "Phys. Rev. Lett.",
      volume         = "44",
      year           = "1980",
      pages          = "912",
      doi            = "10.1103/PhysRevLett.44.912",
      reportNumber   = "MDDP-TR-80-060, MDDP-PP-80-105, CCNY-HEP-79-10",
      SLACcitation   = "%%CITATION = PRLTA,44,912;%%"
}

@article{GellMann:1980vs,
      author         = "Gell-Mann, Murray and Ramond, Pierre and Slansky,
                        Richard",
      title          = "{Complex Spinors and Unified Theories}",
      booktitle      = "{Supergravity Workshop Stony Brook, New York, September
                        27-28, 1979}",
      journal        = "Conf. Proc.",
      volume         = "C790927",
      year           = "1979",
      pages          = "315-321",
      eprint         = "1306.4669",
      archivePrefix  = "arXiv",
      primaryClass   = "hep-th",
      reportNumber   = "PRINT-80-0576",
      SLACcitation   = "%%CITATION = ARXIV:1306.4669;%%"
}

@article{Zee:1980ai,
      author         = "Zee, A.",
      title          = "{A Theory of Lepton Number Violation, Neutrino Majorana
                        Mass, and Oscillation}",
      journal        = "Phys. Lett.",
      volume         = "93B",
      year           = "1980",
      pages          = "389",
      doi            = "10.1016/0370-2693(80)90349-4,
                        10.1016/0370-2693(80)90193-8",
      note           = "[Erratum: Phys. Lett.95B,461(1980)]",
      reportNumber   = "UPR-0150T",
      SLACcitation   = "%%CITATION = PHLTA,93B,389;%%"
}

@article{Babu:1988ki,
      author         = "Babu, K. S.",
      title          = "{Model of 'Calculable' Majorana Neutrino Masses}",
      journal        = "Phys. Lett.",
      volume         = "B203",
      year           = "1988",
      pages          = "132-136",
      doi            = "10.1016/0370-2693(88)91584-5",
      reportNumber   = "UR-1039, ER13065-517",
      SLACcitation   = "%%CITATION = PHLTA,B203,132;%%"
}

@article{ArkaniHamed:1998vp,
      author         = "Arkani-Hamed, Nima and Dimopoulos, Savas and Dvali, G. R.
                        and March-Russell, John",
      title          = "{Neutrino masses from large extra dimensions}",
      booktitle      = "{SUSY 98 Conference Oxford, England, July 11-17, 1998}",
      journal        = "Phys. Rev.",
      volume         = "D65",
      year           = "2001",
      pages          = "024032",
      doi            = "10.1103/PhysRevD.65.024032",
      eprint         = "hep-ph/9811448",
      archivePrefix  = "arXiv",
      primaryClass   = "hep-ph",
      reportNumber   = "SLAC-PUB-8014, SU-ITP-98-64",
      SLACcitation   = "%%CITATION = HEP-PH/9811448;%%"
}

@article{Mohapatra:1986bd,
      author         = "Mohapatra, R. N. and Valle, J. W. F.",
      title          = "{Neutrino Mass and Baryon Number Nonconservation in
                        Superstring Models}",
      booktitle      = "{Sixty years of double beta decay: From nuclear physics
                        to beyond standard model particle physics}",
      journal        = "Phys. Rev.",
      volume         = "D34",
      year           = "1986",
      pages          = "1642",
      doi            = "10.1103/PhysRevD.34.1642",
      reportNumber   = "MdDP-PP-86-127",
      SLACcitation   = "%%CITATION = PHRVA,D34,1642;%%"
}

@article{GonzalezGarcia:1988rw,
      author         = "Gonzalez-Garcia, M. C. and Valle, J. W. F.",
      title          = "{Fast Decaying Neutrinos and Observable Flavor Violation
                        in a New Class of Majoron Models}",
      journal        = "Phys. Lett.",
      volume         = "B216",
      year           = "1989",
      pages          = "360-366",
      doi            = "10.1016/0370-2693(89)91131-3",
      reportNumber   = "CERN-TH-5170-88, FTUV-10-88",
      SLACcitation   = "%%CITATION = PHLTA,B216,360;%%"
}

@article{Malinsky:2005bi,
      author         = "Malinsky, Michal and Romao, J. C. and Valle, J. W. F.",
      title          = "{Novel supersymmetric SO(10) seesaw mechanism}",
      journal        = "Phys. Rev. Lett.",
      volume         = "95",
      year           = "2005",
      pages          = "161801",
      doi            = "10.1103/PhysRevLett.95.161801",
      eprint         = "hep-ph/0506296",
      archivePrefix  = "arXiv",
      primaryClass   = "hep-ph",
      reportNumber   = "IFIC-05-28",
      SLACcitation   = "%%CITATION = HEP-PH/0506296;%%"
}

@article{Mohapatra:2005wg,
      author         = "Mohapatra, R. N. and others",
      title          = "{Theory of neutrinos: A White paper}",
      journal        = "Rept. Prog. Phys.",
      volume         = "70",
      year           = "2007",
      pages          = "1757-1867",
      doi            = "10.1088/0034-4885/70/11/R02",
      eprint         = "hep-ph/0510213",
      archivePrefix  = "arXiv",
      primaryClass   = "hep-ph",
      reportNumber   = "FERMILAB-TM-2342-T, SLAC-PUB-11622",
      SLACcitation   = "%%CITATION = HEP-PH/0510213;%%"
}

@article{Mishra:2019oqq,
    author = "Mishra, Subhasmita and Kumar Behera, Mitesh and Mohanta, Rukmani and Patra, Sudhanwa and Singirala, Shivaramakrishna",
    title = "{Neutrino Phenomenology and Dark matter in an $A_4$ flavour extended B-L model}",
    eprint = "1907.06429",
    archivePrefix = "arXiv",
    primaryClass = "hep-ph",
    doi = "10.1140/epjc/s10052-020-7968-9",
    journal = "Eur. Phys. J. C",
    volume = "80",
    number = "5",
    pages = "420",
    year = "2020"
}

@article{Ma:2015fpa,
    author = "Ma, Ernest",
    title = "{Neutrino mixing: $A_4$ variations}",
    eprint = "1510.02501",
    archivePrefix = "arXiv",
    primaryClass = "hep-ph",
    reportNumber = "UCRHEP-T556-(OCT-2015)",
    doi = "10.1016/j.physletb.2015.11.049",
    journal = "Phys. Lett. B",
    volume = "752",
    pages = "198--200",
    year = "2016"
}

@article{Kobayashi:2018wkl,
    author = "Kobayashi, Tatsuo and Shimizu, Yusuke and Takagi, Kenta and Tanimoto, Morimitsu and Tatsuishi, Takuya H. and Uchida, Hikaru",
    title = "{Finite modular subgroups for fermion mass matrices and baryon/lepton number violation}",
    eprint = "1812.11072",
    archivePrefix = "arXiv",
    primaryClass = "hep-ph",
    reportNumber = "EPHOU-18-017, HUPD1810",
    doi = "10.1016/j.physletb.2019.05.034",
    journal = "Phys. Lett. B",
    volume = "794",
    pages = "114--121",
    year = "2019"
}

@article{Abbas:2020qzc,
    author = "Abbas, Mohamed",
    title = "{Flavor masses and mixing in modular $A$$_{4}$ Symmetry}",
    eprint = "2002.01929",
    archivePrefix = "arXiv",
    primaryClass = "hep-ph",
    month = "2",
    year = "2020"
}

@article{Liu:2020akv,
    author = "Liu, Xiang-Gan and Yao, Chang-Yuan and Ding, Gui-Jun",
    title = "{Modular Invariant Quark and Lepton Models in Double Covering of $S_4$ Modular Group}",
    eprint = "2006.10722",
    archivePrefix = "arXiv",
    primaryClass = "hep-ph",
    reportNumber = "USTC-ICTS/PCFT-20-18",
    month = "6",
    year = "2020"
}

@article{Kang:2026osw,
    author = "Kang, Sin Kyu and Kumar, Ranjeet and Okada, Hiroshi",
    title = "{Radiative Dirac Neutrino Masses from Modular $S_3$ Symmetry in an Axion Model}",
    eprint = "2601.22740",
    archivePrefix = "arXiv",
    primaryClass = "hep-ph",
    month = "1",
    year = "2026"
}

@article{King:2020qaj,
    author = "King, Simon J.D. and King, Stephen F.",
    title = "{Fermion Mass Hierarchies from Modular Symmetry}",
    eprint = "2002.00969",
    archivePrefix = "arXiv",
    primaryClass = "hep-ph",
    month = "2",
    year = "2020"
}

@article{Nomura:2024vus,
    author = "Nomura, Takaaki and Okada, Hiroshi",
    title = "{A More Novel Approach of Radiative Linear Seesaw in a Modular A4 Symmetry}",
    eprint = "2410.21843",
    archivePrefix = "arXiv",
    primaryClass = "hep-ph",
    doi = "10.1093/ptep/ptaf044",
    journal = "PTEP",
    volume = "2025",
    number = "4",
    pages = "043B04",
    year = "2025"
}

@article{deMedeirosVarzielas:2023crv,
    author = "de Medeiros Varzielas, I. and Levy, M. and Penedo, J. T. and Petcov, S. T.",
    title = "{Quarks at the modular S$_{4}$ cusp}",
    eprint = "2307.14410",
    archivePrefix = "arXiv",
    primaryClass = "hep-ph",
    reportNumber = "SISSA 11/2023/FISI, CFTP/23-002",
    doi = "10.1007/JHEP09(2023)196",
    journal = "JHEP",
    volume = "09",
    pages = "196",
    year = "2023"
}

@article{Wang:2019xbo,
    author = "Wang, Xin",
    title = "{Lepton Flavor Mixing and CP Violation in the Minimal Type-(I+II) Seesaw Model with a Modular $A_4$ Symmetry}",
    eprint = "1912.13284",
    archivePrefix = "arXiv",
    primaryClass = "hep-ph",
    month = "12",
    year = "2019"
}

@article{Lu:2019vgm,
    author = "Lu, Jun-Nan and Liu, Xiang-Gan and Ding, Gui-Jun",
    title = "{Modular symmetry origin of texture zeros and quark lepton unification}",
    eprint = "1912.07573",
    archivePrefix = "arXiv",
    primaryClass = "hep-ph",
    reportNumber = "USTC-ICTS-19-33",
    doi = "10.1103/PhysRevD.101.115020",
    journal = "Phys. Rev. D",
    volume = "101",
    number = "11",
    pages = "115020",
    year = "2020"
}

@inbook{Feruglio:2017spp,
    author = "Feruglio, Ferruccio",
    archivePrefix = "arXiv",
    booktitle = "{From My Vast Repertoire ...}: {Guido Altarelli's Legacy}",
    doi = "10.1142/9789813238053\_0012",
    eprint = "1706.08749",
    pages = "227--266",
    primaryClass = "hep-ph",
    reportNumber = "DFPD-2017-TH-09",
    title = "{Are neutrino masses modular forms?}",
    year = "2019"
}

@article{article,
author = {Vicente, Avelino},
year = {2011},
month = {04},
pages = {},
title = {Phenomenology of supersymmetric neutrino mass models}
}

@article{Kobayashi:2019gtp,
    author = "Kobayashi, Tatsuo and Nomura, Takaaki and Shimomura, Takashi",
    archivePrefix = "arXiv",
    eprint = "1912.00637",
    month = "12",
    primaryClass = "hep-ph",
    reportNumber = "EPHOU-19-018, KIAS-P19068, UME-PP-011",
    title = "{Type II seesaw models with modular $A_4$ symmetry}",
    year = "2019"
}

@article{deMedeirosVarzielas:2022ihu,
    author = "de Medeiros Varzielas, Ivo and Louren{\c{c}}o, Jo{\~a}o",
    title = "{Two A5 modular symmetries for Golden Ratio 2 mixing}",
    eprint = "2206.14869",
    archivePrefix = "arXiv",
    primaryClass = "hep-ph",
    doi = "10.1016/j.nuclphysb.2022.115974",
    journal = "Nucl. Phys. B",
    volume = "984",
    pages = "115974",
    year = "2022"
}

@article{Yao:2020zml,
    author = "Yao, Chang-Yuan and Liu, Xiang-Gan and Ding, Gui-Jun",
    title = "{Fermion masses and mixing from the double cover and metaplectic cover of the $A_5$ modular group}",
    eprint = "2011.03501",
    archivePrefix = "arXiv",
    primaryClass = "hep-ph",
    reportNumber = "USTC-ICTS/PCFT-20-36",
    doi = "10.1103/PhysRevD.103.095013",
    journal = "Phys. Rev. D",
    volume = "103",
    number = "9",
    pages = "095013",
    year = "2021"
}

@article{Nomura:2019xsb,
    author = "Nomura, Takaaki and Okada, Hiroshi and Patra, Sudhanwa",
    archivePrefix = "arXiv",
    eprint = "1912.00379",
    month = "12",
    primaryClass = "hep-ph",
    reportNumber = "KIAS-P19067, APCTP Pre2019 - 026",
    title = "{An Inverse Seesaw model with $A_4$-modular symmetry}",
    year = "2019"
}

@software{FlavorPy,
      author        = {Baur, Alexander},
      title         = "{FlavorPy}",
      year          = {2024},
      publisher     = {Zenodo},
      version       = {v0.1.0},
      doi           = {10.5281/zenodo.11060597},
      url           = "\url{https://doi.org/10.5281/zenodo.11060597}"
    }

@techreport{Gallier2010Schur,
  author      = {Jean H. Gallier},
  title       = {The Schur Complement and Symmetric Positive Semidefinite (and Definite) Matrices},
  institution = {University of Pennsylvania},
  year        = {2010},
  url         = {https://www.cis.upenn.edu/~jean/schur-comp.pdf}
}

@article{CarcamoHernandez:2019pmy,
    author = "Cárcamo Hernández, A.E. and Marchant González, Juan and Saldaña-Salazar, U.J.",
    title = "{Viable low-scale model with universal and inverse seesaw mechanisms}",
    eprint = "1904.09993",
    archivePrefix = "arXiv",
    primaryClass = "hep-ph",
    doi = "10.1103/PhysRevD.100.035024",
    journal = "Phys. Rev. D",
    volume = "100",
    number = "3",
    pages = "035024",
    year = "2019"
}

@article{Novichkov:2018ovf,
    author = "Novichkov, P.P. and Penedo, J.T. and Petcov, S.T. and Titov, A.V.",
    title = "{Modular S$_{4}$ models of lepton masses and mixing}",
    eprint = "1811.04933",
    archivePrefix = "arXiv",
    primaryClass = "hep-ph",
    reportNumber = "SISSA 47/2018/FISI, IPMU18-0187, IPPP/18/98",
    doi = "10.1007/JHEP04(2019)005",
    journal = "JHEP",
    volume = "04",
    pages = "005",
    year = "2019"
}

@article{Gui-JunDing:2019wap,
    author = "Ding, Gui-Jun and King, Stephen F. and Liu, Xiang-Gan and Lu, Jun-Nan",
    title = "{Modular S$_{4}$ and A$_{4}$ symmetries and their fixed points: new predictive examples of lepton mixing}",
    eprint = "1910.03460",
    archivePrefix = "arXiv",
    primaryClass = "hep-ph",
    reportNumber = "USTC-ICTS-19-26",
    doi = "10.1007/JHEP12(2019)030",
    journal = "JHEP",
    volume = "12",
    pages = "030",
    year = "2019"
}

@article{Ding:2019zxk,
    author = "Ding, Gui-Jun and King, Stephen F. and Liu, Xiang-Gan",
    title = "{Modular A$_{4}$ symmetry models of neutrinos and charged leptons}",
    eprint = "1907.11714",
    archivePrefix = "arXiv",
    primaryClass = "hep-ph",
    reportNumber = "USTC-ICTS-19-16",
    doi = "10.1007/JHEP09(2019)074",
    journal = "JHEP",
    volume = "09",
    pages = "074",
    year = "2019"
}

@article{Asaka:2019vev,
    author = "Asaka, Takehiko and Heo, Yongtae and Tatsuishi, Takuya H. and Yoshida, Takahiro",
    title = "{Modular $A_4$ invariance and leptogenesis}",
    eprint = "1909.06520",
    archivePrefix = "arXiv",
    primaryClass = "hep-ph",
    reportNumber = "EPHOU-19-013",
    doi = "10.1007/JHEP01(2020)144",
    journal = "JHEP",
    volume = "01",
    pages = "144",
    year = "2020"
}

@article{Nomura:2023usj,
    author = "Nomura, Takaaki and Okada, Hiroshi and Otsuka, Hajime",
    title = "{Texture zeros realization in a three-loop radiative neutrino mass model from modular A4 symmetry}",
    eprint = "2309.13921",
    archivePrefix = "arXiv",
    primaryClass = "hep-ph",
    reportNumber = "KYUSHU-HET-268",
    doi = "10.1016/j.nuclphysb.2024.116579",
    journal = "Nucl. Phys. B",
    volume = "1004",
    pages = "116579",
    year = "2024"
}

@article{RickyDevi:2024ijc,
    author = "Ricky Devi, Maibam",
    title = "{Neutrino Masses and Higher Degree Siegel Modular Forms}",
    eprint = "2401.16257",
    archivePrefix = "arXiv",
    primaryClass = "hep-ph",
    month = "1",
    year = "2024"
}

@article{Novichkov:2018nkm,
    author = "Novichkov, P.P. and Penedo, J.T. and Petcov, S.T. and Titov, A.V.",
    title = "{Modular A$_{5}$ symmetry for flavour model building}",
    eprint = "1812.02158",
    archivePrefix = "arXiv",
    primaryClass = "hep-ph",
    reportNumber = "SISSA 54/2018/FISI, IPMU18-0202, IPPP/18/105",
    doi = "10.1007/JHEP04(2019)174",
    journal = "JHEP",
    volume = "04",
    pages = "174",
    year = "2019"
}

@article{Aker:2019uuj,
    author = "Aker, M. and others",
    collaboration = "KATRIN",
    title = "{Improved Upper Limit on the Neutrino Mass from a Direct Kinematic Method by KATRIN}",
    eprint = "1909.06048",
    archivePrefix = "arXiv",
    primaryClass = "hep-ex",
    doi = "10.1103/PhysRevLett.123.221802",
    journal = "Phys. Rev. Lett.",
    volume = "123",
    number = "22",
    pages = "221802",
    year = "2019"
}

@article{Ishimori:2010au,
    author = "Ishimori, Hajime and Kobayashi, Tatsuo and Ohki, Hiroshi and Shimizu, Yusuke and Okada, Hiroshi and Tanimoto, Morimitsu",
    title = "{Non-Abelian Discrete Symmetries in Particle Physics}",
    eprint = "1003.3552",
    archivePrefix = "arXiv",
    primaryClass = "hep-th",
    reportNumber = "KUNS-2260",
    doi = "10.1143/PTPS.183.1",
    journal = "Prog. Theor. Phys. Suppl.",
    volume = "183",
    pages = "1--163",
    year = "2010"
}

@article{Alduino:2017ehq,
    author = "Alduino, C. and others",
    collaboration = "CUORE",
    title = "{First Results from CUORE: A Search for Lepton Number Violation via $0\nu\beta\beta$ Decay of $^{130}$Te}",
    eprint = "1710.07988",
    archivePrefix = "arXiv",
    primaryClass = "nucl-ex",
    doi = "10.1103/PhysRevLett.120.132501",
    journal = "Phys. Rev. Lett.",
    volume = "120",
    number = "13",
    pages = "132501",
    year = "2018"
}

@article{Agostini:2019hzm,
    author = "Agostini, M. and others",
    collaboration = "GERDA",
    title = "{Probing Majorana neutrinos with double-$\beta$ decay}",
    eprint = "1909.02726",
    archivePrefix = "arXiv",
    primaryClass = "hep-ex",
    doi = "10.1126/science.aav8613",
    journal = "Science",
    volume = "365",
    pages = "1445",
    year = "2019"
}

@article{Novichkov:2020eep,
    author = "Novichkov, P. P. and Penedo, J. T. and Petcov, S. T.",
    title = "{Double cover of modular $S_4$ for flavour model building}",
    eprint = "2006.03058",
    archivePrefix = "arXiv",
    primaryClass = "hep-ph",
    reportNumber = "SISSA 12/2020/FISI, IPMU20-0065, CFTP/20-006",
    doi = "10.1016/j.nuclphysb.2020.115301",
    journal = "Nucl. Phys. B",
    volume = "963",
    pages = "115301",
    year = "2021"
}

@article{Chauhan:2023faf,
    author = "Chauhan, Garv and Dev, P. S. Bhupal and Dubovyk, Ievgen and Dziewit, Bartosz and Flieger, Wojciech and Grzanka, Krzysztof and Gluza, Janusz and Karmakar, Biswajit and Zi{\k{e}}ba, Szymon",
    title = "{Phenomenology of lepton masses and mixing with discrete flavor symmetries}",
    eprint = "2310.20681",
    archivePrefix = "arXiv",
    primaryClass = "hep-ph",
    doi = "10.1016/j.ppnp.2024.104126",
    journal = "Prog. Part. Nucl. Phys.",
    volume = "138",
    pages = "104126",
    year = "2024"
}

@article{Qu:2024rns,
    author = "Qu, Bu-Yao and Ding, Gui-Jun",
    title = "{Non-holomorphic modular flavor symmetry}",
    eprint = "2406.02527",
    archivePrefix = "arXiv",
    primaryClass = "hep-ph",
    doi = "10.1007/JHEP08(2024)136",
    journal = "JHEP",
    volume = "08",
    pages = "136",
    year = "2024"
}

@article{Mishra:2023ekx,
    author = "Mishra, Priya and Behera, Mitesh Kumar and Panda, Papia and Ghosh, Monojit and Mohanta, Rukmani",
    title = "{Exploring models with modular symmetry in neutrino oscillation experiments}",
    eprint = "2305.08576",
    archivePrefix = "arXiv",
    primaryClass = "hep-ph",
    doi = "10.1007/JHEP09(2023)144",
    journal = "JHEP",
    volume = "09",
    pages = "144",
    year = "2023"
}

@article{Kashav:2021zir,
    author = "Kashav, Monal and Verma, Surender",
    title = "{Broken Scaling Neutrino Mass Matrix and Leptogenesis based on A$_4$ Modular invariance}",
    eprint = "2103.07207",
    archivePrefix = "arXiv",
    primaryClass = "hep-ph",
    month = "3",
    year = "2021"
}

@article{DUNE:2021cuw,
    author = "Abi, B. and others",
    collaboration = "DUNE",
    title = "{Experiment Simulation Configurations Approximating DUNE TDR}",
    eprint = "2103.04797",
    archivePrefix = "arXiv",
    primaryClass = "hep-ex",
    reportNumber = "FERMILAB-FN-1125-ND",
    month = "3",
    year = "2021"
}

@article{Ma:2001dn,
      author         = "Ma, Ernest and Rajasekaran, G.",
      title          = "{Softly broken A(4) symmetry for nearly degenerate
                        neutrino masses}",
      journal        = "Phys. Rev.",
      volume         = "D64",
      year           = "2001",
      pages          = "113012",
      doi            = "10.1103/PhysRevD.64.113012",
      eprint         = "hep-ph/0106291",
      archivePrefix  = "arXiv",
      primaryClass   = "hep-ph",
      reportNumber   = "UCRHEP-T308",
      SLACcitation   = "%%CITATION = HEP-PH/0106291;%%"
}

@article{Pontecorvo:1967fh,
    author = "Pontecorvo, B.",
    title = "{Neutrino Experiments and the Problem of Conservation of Leptonic Charge}",
    journal = "Sov. Phys. JETP",
    volume = "26",
    pages = "984--988",
    year = "1968"
}

@article{Faessler:2020sgs,
    author = "Faessler, Amand",
    title = "{Status of the determination of the electron--neutrino mass}",
    doi = "10.1016/j.ppnp.2020.103789",
    journal = "Prog. Part. Nucl. Phys.",
    volume = "113",
    pages = "103789",
    year = "2020"
}

@article{Penedo:2018nmg,
    author = "Penedo, J.T. and Petcov, S.T.",
    archivePrefix = "arXiv",
    doi = "10.1016/j.nuclphysb.2018.12.016",
    eprint = "1806.11040",
    journal = "Nucl.\ Phys.\ B",
    pages = "292--307",
    primaryClass = "hep-ph",
    reportNumber = "SISSA 25/2018/FISI, IPMU18-0121, SISSA-25-2018-FISI",
    title = "{Lepton Masses and Mixing from Modular $S_4$ Symmetry}",
    volume = "939",
    year = "2019"
}

@article{Kobayashi:2019xvz,
    author = "Kobayashi, Tatsuo and Shimizu, Yusuke and Takagi, Kenta and Tanimoto, Morimitsu and Tatsuishi, Takuya H.",
    title = "{$A_4$ lepton flavor model and modulus stabilization from $S_4$ modular symmetry}",
    eprint = "1909.05139",
    archivePrefix = "arXiv",
    primaryClass = "hep-ph",
    reportNumber = "EPHOU-19-012, HUPD1912",
    doi = "10.1103/PhysRevD.100.115045",
    journal = "Phys. Rev. D",
    volume = "100",
    number = "11",
    pages = "115045",
    year = "2019",
    note = "[Erratum: Phys.Rev.D 101, 039904 (2020)]"
}

@article{Giuliani:2019uno,
    author = {Giuliani, A. and Gomez Cadenas, J. J. and Pascoli, S. and Previtali, E. and Saakyan, R. and Sch\"affner, K. and Sch\"onert, S.},
    collaboration = "APPEC Committee",
    title = "{Double Beta Decay APPEC Committee Report}",
    eprint = "1910.04688",
    archivePrefix = "arXiv",
    primaryClass = "hep-ex",
    month = "10",
    year = "2019"
}

@article{KamLAND-Zen:2024eml,
    author = "Abe, S. and others",
    collaboration = "KamLAND-Zen",
    title = "{Search for Majorana Neutrinos with the Complete KamLAND-Zen Dataset}",
    eprint = "2406.11438",
    archivePrefix = "arXiv",
    primaryClass = "hep-ex",
    month = "6",
    year = "2024"
}

@article{LEGEND:2021bnm,
    author = "Abgrall, N. and others",
    collaboration = "LEGEND",
    title = "{The Large Enriched Germanium Experiment for Neutrinoless $\beta\beta$ Decay}: {LEGEND-1000 Preconceptual Design Report}",
    eprint = "2107.11462",
    archivePrefix = "arXiv",
    primaryClass = "physics.ins-det",
    month = "7",
    year = "2021"
}

@article{nEXO:2021ujk,
    author = "Adhikari, G. and others",
    collaboration = "nEXO",
    title = "{nEXO: neutrinoless double beta decay search beyond 10$^{28}$ year half-life sensitivity}",
    eprint = "2106.16243",
    archivePrefix = "arXiv",
    primaryClass = "nucl-ex",
    doi = "10.1088/1361-6471/ac3631",
    journal = "J. Phys. G",
    volume = "49",
    number = "1",
    pages = "015104",
    year = "2022"
}

@inproceedings{Project8:2022wqh,
    author = "Esfahani, A. Ashtari and others",
    collaboration = "Project 8",
    title = "{The Project 8 Neutrino Mass Experiment}",
    booktitle = "{Snowmass 2021}",
    eprint = "2203.07349",
    archivePrefix = "arXiv",
    primaryClass = "nucl-ex",
    month = "3",
    year = "2022"
}

@article{DESI:2024mwx,
    author = "Adame, A. G. and others",
    collaboration = "DESI",
    title = "{DESI 2024 VI: cosmological constraints from the measurements of baryon acoustic oscillations}",
    eprint = "2404.03002",
    archivePrefix = "arXiv",
    primaryClass = "astro-ph.CO",
    reportNumber = "FERMILAB-PUB-24-0154-PPD",
    doi = "10.1088/1475-7516/2025/02/021",
    journal = "JCAP",
    volume = "02",
    pages = "021",
    year = "2025"
}

@article{Euclid:2024imf,
    author = "Archidiacono, M. and others",
    collaboration = "Euclid",
    title = "{Euclid preparation - LIV. Sensitivity to neutrino parameters}",
    eprint = "2405.06047",
    archivePrefix = "arXiv",
    primaryClass = "astro-ph.CO",
    reportNumber = "TTK-24-17",
    doi = "10.1051/0004-6361/202450859",
    journal = "Astron. Astrophys.",
    volume = "693",
    pages = "A58",
    year = "2025"
}

@book{Ishimori:2012zz,
    author = "Ishimori, Hajime and Kobayashi, Tatsuo and Ohki, Hiroshi and Okada, Hiroshi and Shimizu, Yusuke and Tanimoto, Morimitsu",
    title = "{An introduction to non-Abelian discrete symmetries for particle physicists}",
    doi = "10.1007/978-3-642-30805-5",
    volume = "858",
    year = "2012"
}

@article{Tapender:2026ets,
    author = "Tapender and Verma, Surender",
    title = "{Tri-Resonant Leptogenesis in a Non-Holomorphic Modular A$_4$ Scotogenic Model}",
    eprint = "2602.17243",
    archivePrefix = "arXiv",
    primaryClass = "hep-ph",
    month = "2",
    year = "2026"
}

@article{Nasri:2026nbf,
    author = "Nasri, Salah and Singh, Labh and Tapender and Verma, Surender",
    title = "{Dark-Portal Leptogenesis in a Non-Holomorphic Modular Scoto-Seesaw Model}",
    eprint = "2601.06435",
    archivePrefix = "arXiv",
    primaryClass = "hep-ph",
    month = "1",
    year = "2026"
}

@article{Priya:2025wdm,
    author = "Priya and Singh, Labh and Chauhan, B. C. and Verma, Surender",
    title = "{Type-III seesaw in non-holomorphic modular symmetry and leptogenesis}",
    eprint = "2508.05047",
    archivePrefix = "arXiv",
    primaryClass = "hep-ph",
    doi = "10.1007/JHEP01(2026)036",
    journal = "JHEP",
    volume = "01",
    pages = "036",
    year = "2026"
}

@article{Nomura:2024atp,
    author = "Nomura, Takaaki and Okada, Hiroshi",
    title = "{Type-II seesaw of a non-holomorphic modular A4 symmetry}",
    eprint = "2408.01143",
    archivePrefix = "arXiv",
    primaryClass = "hep-ph",
    doi = "10.1016/j.physletb.2025.139763",
    journal = "Phys. Lett. B",
    volume = "868",
    pages = "139763",
    year = "2025"
}

@article{Nomura:2024vzw,
    author = "Nomura, Takaaki and Okada, Hiroshi and Popov, Oleg",
    title = "{Non-holomorphic modular A4 symmetric scotogenic model}",
    eprint = "2409.12547",
    archivePrefix = "arXiv",
    primaryClass = "hep-ph",
    doi = "10.1016/j.physletb.2024.139171",
    journal = "Phys. Lett. B",
    volume = "860",
    pages = "139171",
    year = "2025"
}

@article{Nomura:2024nwh,
    author = "Nomura, Takaaki and Okada, Hiroshi",
    title = "{Zee model in a non-holomorphic modular A4 symmetry}",
    eprint = "2412.18095",
    archivePrefix = "arXiv",
    primaryClass = "hep-ph",
    doi = "10.1016/j.physletb.2025.139618",
    journal = "Phys. Lett. B",
    volume = "867",
    pages = "139618",
    year = "2025"
}

@article{Nomura:2025ovm,
    author = "Nomura, Takaaki and Okada, Hiroshi and Wang, Xing-Yu",
    title = "{A radiative neutrino mass model with leptoquarks under non-holomorphic modular A$_{4}$ symmetry}",
    eprint = "2504.21404",
    archivePrefix = "arXiv",
    primaryClass = "hep-ph",
    doi = "10.1007/JHEP09(2025)163",
    journal = "JHEP",
    volume = "09",
    pages = "163",
    year = "2025"
}

@article{Nomura:2025raf,
    author = "Nomura, Takaaki and Okada, Hiroshi",
    title = "{Neutrino mass model at a three-loop level from a non-holomorphic modular $A_4$ symmetry}",
    eprint = "2506.02639",
    archivePrefix = "arXiv",
    primaryClass = "hep-ph",
    month = "6",
    year = "2025"
}

@article{Nomura:2025bph,
    author = "Nomura, Takaaki and Okada, Hiroshi",
    title = "{A new type of lepton seesaw model in a modular $A_4$ symmetry}",
    eprint = "2503.19251",
    archivePrefix = "arXiv",
    primaryClass = "hep-ph",
    month = "3",
    year = "2025"
}

@article{Kang:2024jnp,
    author = "Kang, Sin Kyu and Okada, Hiroshi",
    title = "{Neutrino masses and mixing in an axion model}",
    eprint = "2408.14942",
    archivePrefix = "arXiv",
    primaryClass = "hep-ph",
    doi = "10.1140/epjc/s10052-025-14636-6",
    journal = "Eur. Phys. J. C",
    volume = "85",
    number = "8",
    pages = "917",
    year = "2025"
}

@article{Ding:2024inn,
    author = "Ding, Gui-Jun and Lu, Jun-Nan and Petcov, S. T. and Qu, Bu-Yao",
    title = "{Non-holomorphic modular S$_{4}$ lepton flavour models}",
    eprint = "2408.15988",
    archivePrefix = "arXiv",
    primaryClass = "hep-ph",
    reportNumber = "SISSA 17/2024/FISI",
    doi = "10.1007/JHEP01(2025)191",
    journal = "JHEP",
    volume = "01",
    pages = "191",
    year = "2025"
}

@article{Li:2024svh,
    author = "Li, Cai-Chang and Lu, Jun-Nan and Ding, Gui-Jun",
    title = "{Non-holomorphic modular A$_{5}$ symmetry for lepton masses and mixing}",
    eprint = "2410.24103",
    archivePrefix = "arXiv",
    primaryClass = "hep-ph",
    doi = "10.1007/JHEP12(2024)189",
    journal = "JHEP",
    volume = "12",
    pages = "189",
    year = "2024"
}

@article{Okada:2025jjo,
    author = "Okada, Hiroshi and Orikasa, Yuta",
    title = "{A radiative seesaw in a non-holomorphic modular $S_3$ flavor symmetry}",
    eprint = "2501.15748",
    archivePrefix = "arXiv",
    primaryClass = "hep-ph",
    month = "1",
    year = "2025"
}

@article{Kobayashi:2025hnc,
    author = "Kobayashi, Tatsuo and Okada, Hiroshi and Orikasa, Yuta",
    title = "{Zee-Babu model in a non-holomorphic modular $A_4$ symmetry and modular stabilization}",
    eprint = "2502.12662",
    archivePrefix = "arXiv",
    primaryClass = "hep-ph",
    reportNumber = "EPHOU-25-001",
    month = "2",
    year = "2025"
}

@article{Loualidi:2025tgw,
    author = "Loualidi, Mohamed Amin and Miskaoui, Mohamed and Nasri, Salah",
    title = "{Nonholomorphic A4 modular invariance for fermion masses and mixing in SU(5) GUT}",
    eprint = "2503.12594",
    archivePrefix = "arXiv",
    primaryClass = "hep-ph",
    doi = "10.1103/1py2-cmfx",
    journal = "Phys. Rev. D",
    volume = "112",
    number = "1",
    pages = "015008",
    year = "2025"
}

@article{Nomura:2024ctl,
    author = "Nomura, Takaaki and Okada, Hiroshi",
    title = "{Lepton seesaw model in a modular $A_4$ symmetry}",
    eprint = "2409.10912",
    archivePrefix = "arXiv",
    primaryClass = "hep-ph",
    month = "9",
    year = "2024"
}

@article{Abbas:2025nlv,
    author = "Abbas, Mohammed",
    title = "{Lepton Masses and Mixing in Nonholomorphic Modular A4 Symmetry}",
    doi = "10.31526/PHEP.2025.07",
    journal = "PHEP",
    volume = "2025",
    pages = "7",
    year = "2025"
}

@article{Li:2025kcr,
    author = "Li, Cai-Chang and Ding, Gui-Jun",
    title = "{Lepton models from non-holomorphic $A^{\prime}_{5}$ modular flavor symmetry}",
    eprint = "2509.15183",
    archivePrefix = "arXiv",
    primaryClass = "hep-ph",
    month = "9",
    year = "2025"
}

@article{Dey:2025zld,
    author = "Dey, Manash",
    title = "{The Seesaw Evaded Modular Dirac Framework}",
    eprint = "2509.10373",
    archivePrefix = "arXiv",
    primaryClass = "hep-ph",
    month = "9",
    year = "2025"
}

@article{Nanda:2025lem,
    author = "Nanda, Swaraj Kumar and Ricky Devi, Maibam and Patra, Sudhanwa",
    title = "{Non-Holomorphic $A_4$ Modular Symmetry in Type-I Seesaw: Implications for Neutrino Masses and Leptogenesis}",
    eprint = "2509.22108",
    archivePrefix = "arXiv",
    primaryClass = "hep-ph",
    month = "9",
    year = "2025"
}

@article{Kumar:2025bfe,
    author = "Kumar, Bhabana and Das, Mrinal Kumar",
    title = "{Leptogenesis, 0{\ensuremath{\nu}}{\ensuremath{\beta}}{\ensuremath{\beta}} and lepton flavor violation in modular left-right asymmetric model with polyharmonic Maa{\ss} forms}",
    eprint = "2504.21701",
    archivePrefix = "arXiv",
    primaryClass = "hep-ph",
    doi = "10.1007/JHEP09(2025)071",
    journal = "JHEP",
    volume = "09",
    pages = "071",
    year = "2025"
}

@article{Behera:2025tpj,
    author = "Behera, Mitesh Kumar and Ittisamai, Pawin and Pongkitivanichkul, Chakrit and Uttayarat, Patipan",
    title = "{Phenomenology of inverse seesaw using $S_3$ modular symmetry}",
    eprint = "2504.12954",
    archivePrefix = "arXiv",
    primaryClass = "hep-ph",
    doi = "10.1140/epjc/s10052-025-15017-9",
    journal = "Eur. Phys. J. C",
    volume = "85",
    number = "11",
    pages = "1316",
    year = "2025"
}

@article{Mishra:2020gxg,
    author = "Mishra, Subhasmita",
    title = "{Neutrino mixing and Leptogenesis with modular $S_3$ symmetry in the framework of type III seesaw}",
    eprint = "2008.02095",
    archivePrefix = "arXiv",
    primaryClass = "hep-ph",
    month = "8",
    year = "2020"
}

@article{Meloni:2023aru,
    author = "Meloni, Davide and Parriciatu, Matteo",
    title = "{A simplest modular S$_{3}$ model for leptons}",
    eprint = "2306.09028",
    archivePrefix = "arXiv",
    primaryClass = "hep-ph",
    doi = "10.1007/JHEP09(2023)043",
    journal = "JHEP",
    volume = "09",
    pages = "043",
    year = "2023"
}

@article{Behera:2024ark,
    author = "Behera, Mitesh Kumar and Ittisamai, Pawin and Pongkitivanichkul, Chakrit and Uttayarat, Patipan",
    title = "{Neutrino phenomenology in the modular S3 seesaw model}",
    eprint = "2403.00593",
    archivePrefix = "arXiv",
    primaryClass = "hep-ph",
    doi = "10.1103/PhysRevD.110.035004",
    journal = "Phys. Rev. D",
    volume = "110",
    number = "3",
    pages = "035004",
    year = "2024"
}

@article{Marciano:2024nwm,
    author = "Marciano, Simone and Meloni, Davide and Parriciatu, Matteo",
    title = "{Minimal seesaw and leptogenesis with the smallest modular finite group}",
    eprint = "2402.18547",
    archivePrefix = "arXiv",
    primaryClass = "hep-ph",
    doi = "10.1007/JHEP05(2024)020",
    journal = "JHEP",
    volume = "05",
    pages = "020",
    year = "2024"
}

@article{Belfkir:2024wdn,
    author = "Belfkir, Mohamed and Loualidi, Mohamed Amin and Nasri, Salah",
    title = "{Fermion Masses and Mixing in Pati{\textendash}Salam Unification with S3 Modular Symmetry}",
    eprint = "2501.00302",
    archivePrefix = "arXiv",
    primaryClass = "hep-ph",
    doi = "10.1093/ptep/ptaf032",
    journal = "PTEP",
    volume = "2025",
    number = "3",
    pages = "033B05",
    year = "2025"
}

@article{Okada:2019xqk,
    author = "Okada, Hiroshi and Orikasa, Yuta",
    title = "{Modular $S_3$ symmetric radiative seesaw model}",
    eprint = "1907.04716",
    archivePrefix = "arXiv",
    primaryClass = "hep-ph",
    reportNumber = "APCTP Pre2019-017",
    doi = "10.1103/PhysRevD.100.115037",
    journal = "Phys. Rev. D",
    volume = "100",
    number = "11",
    pages = "115037",
    year = "2019"
}

@article{Ding:2021zbg,
    author = "Ding, Gui-Jun and King, Stephen F. and Yao, Chang-Yuan",
    title = "{Modular $S_4\times SU(5)$ GUT}",
    eprint = "2103.16311",
    archivePrefix = "arXiv",
    primaryClass = "hep-ph",
    reportNumber = "USTC-ICTS/PCFT-21-13",
    doi = "10.1103/PhysRevD.104.055034",
    journal = "Phys. Rev. D",
    volume = "104",
    number = "5",
    pages = "055034",
    year = "2021"
}

@article{King:2019vhv,
    author = "King, Stephen F. and Zhou, Ye-Ling",
    title = "{Trimaximal TM$_1$ mixing with two modular $S_4$ groups}",
    eprint = "1908.02770",
    archivePrefix = "arXiv",
    primaryClass = "hep-ph",
    doi = "10.1103/PhysRevD.101.015001",
    journal = "Phys. Rev. D",
    volume = "101",
    number = "1",
    pages = "015001",
    year = "2020"
}

@article{Behera:2021eut,
    author = "Behera, Mitesh Kumar and Mohanta, Rukmani",
    title = "{Inverse seesaw in $A_5^\prime$ modular symmetry}",
    eprint = "2108.01059",
    archivePrefix = "arXiv",
    primaryClass = "hep-ph",
    doi = "10.1088/1361-6471/ac4d7a",
    journal = "J. Phys. G",
    volume = "49",
    number = "4",
    pages = "045001",
    year = "2022"
}
\end{document}